\documentclass[twocolumn,english]{IEEEtran}
\usepackage[T1]{fontenc}
\usepackage{babel}
\usepackage{array}
\usepackage{booktabs}
\usepackage{multirow}
\usepackage{amsmath}
\usepackage{amssymb}
\usepackage{graphicx}
\makeatletter

\providecommand{\tabularnewline}{\\}

\ifCLASSOPTIONcompsoc
\usepackage[caption=false,font=normalsize,labelfont=sf,textfont=sf]{subfig}
\else
\usepackage[caption=false,font=footnotesize]{subfig}
\fi
\usepackage{cite}
\usepackage{MnSymbol}

\makeatother

\begin{document}

\title{Precision--Energy--Throughput Scaling Of Generic Matrix Multiplication
and Convolution Kernels Via Linear Projections}

\author{Mohammad Ashraful Anam, Paul N. Whatmough and~Yiannis Andreopoulos$^{*}$%
\thanks{\textsuperscript{{*}}Corresponding author. M. A. A. and Y. A. are
with the Electronic and Electrical Engineering Department, University
College London, Roberts Building, Torrington Place, London, WC1E 7JE,
U.K.; Tel. +44 20 7679 7303, Fax. +44 20 7388 9325 (both authors),
Email: \{mohammad.anam.10, i.andreopoulos\}@ucl.ac.uk. P. N. W. is
with ARM Ltd., Cambridge CB1 9NJ, U.K; Email: paul.whatmough@arm.com.
M. A. Anam is a Commonwealth Scholar, funded by the UK government.  This
work was supported in part by U.K. EPSRC, grant EP/M00113X/1. 

This paper appears in the IEEE Trans. on Circuits and Systems for Video Technology, vol. 24, no. 11, pp. 1860-1873, Nov. 2014. Copyright (c) 2014 IEEE. Personal use of this material is permitted.
However, permission to use this material for any other purposes must
be obtained from the IEEE by sending a request to pubs-permissions@ieee.org.%
}}
\maketitle
\begin{abstract}
Generic matrix multiplication (GEMM) and one-dimensional convolution/cross-correlation
(CONV) kernels often constitute the bulk of the compute- and memory-intensive
processing within image/audio recognition and matching systems. We
propose a novel method to scale the \emph{energy} and \emph{processing
throughput} of GEMM and CONV kernels for such error-tolerant multimedia
applications by adjusting the \emph{precision} of computation. Our
technique employs linear projections to the input matrix or signal
data during the top-level GEMM and CONV blocking and reordering. The
GEMM and CONV kernel processing then uses the projected inputs and
the results are accumulated to form the final outputs. Throughput
and energy scaling takes place by changing the number of projections
computed by each kernel, which in turn produces approximate results,
i.e. changes the precision of the performed computation. Results derived
from a voltage- and frequency-scaled ARM Cortex A15 processor running
face recognition and music matching algorithms demonstrate that the
proposed approach allows for $280\%\sim440\%$ increase of processing
throughput and $75\%\sim80\%$ decrease of energy consumption against
optimized GEMM and CONV kernels without any impact in the obtained
recognition or matching accuracy. Even higher gains can be obtained
if one is willing to tolerate some reduction in the accuracy of the
recognition and matching applications. \end{abstract}

\section{Introduction}

\IEEEPARstart{E}{rror-tolerant} multimedia processing \cite{111}
comprises any system that: \emph{(i)} processes large volumes of input
data (image pixels, sensor measurements, database entries, etc.) with
performance-critical digital signal processing (DSP) or linear algebra
kernels (filtering, decomposition, factorization, feature extraction,
principal components, probability mixtures, Monte-Carlo methods, etc.)
and \emph{(ii)} the quality of its results is evaluated in terms of
minimum mean-squared error (MSE) or maximum learning, recognition
or matching rate against ground-truth or training data, rather than
performance bounds for individual inputs. Examples of such \emph{error-tolerant}
(ET) \emph{systems} include: lossy image/video/audio compression \cite{12,26},
computer graphics \cite{15,16}, webpage indexing and retrieval \cite{13},
object and face recognition in video \cite{14,20}, image/video/music
matching \cite{66,63,61,54}, etc. For instance, all face recognition
and webpage ranking algorithms optimize for the expected recall percentage
against ground-truth results and \emph{not} for the worst-case. This
is also because typical input data streams comprise noisy entries
originating from audio/visual sensors, web-crawlers, field-measurement
microsensors, etc. Therefore, ET applications have to tolerate approximations
in their results, and can use this fact to reduce computation time
or energy consumption \cite{111}. 

Two of the most critical linear algebra and DSP kernels used in ET
applications are the generic matrix multiplication (GEMM) and one-dimensional
convolution/cross-correlation (CONV) kernels. This paper proposes
a new approach to systematically scale the computation time and energy
consumption of optimized GEMM and CONV kernels within ET applications
with minimal or no effect in their results.

\subsection{Previous Work }

Several papers have studied techniques to trade-off approximation
versus implementation complexity in GEMM and CONV computations within
\emph{special-purpose systems}. Starting with theory-inspired approaches
for approximate GEMM and CONV kernel realization, Monte-Carlo algorithms
have been proposed for fast approximate matrix multiplication suitable
for massive dataset processing on networked computing systems (aka
\textquotedblleft{}Big Data\textquotedblright{} systems) \cite{124},
such as Google MapReduce and Microsoft Dryad. The concepts of approximate
and stochastic computation in custom hardware were proposed as a means
to achieve complexity-distortion scaling in sum-of-products computations
\cite{27}. Approximate convolution operations in conjunction with
voltage overscaling in custom hardware was proposed recently within
the framework of stochastic computation \cite{8}. 

Other works focus on performance vs. precision tradeoffs of GEMM and
CONV kernels within \emph{specific algorithms}. For example, Merhav
and Kresch \cite{117} presented a novel method of approximate convolution
using discrete cosine transform (DCT) coefficients, which is appropriate
only for DCT-domain processing. Chen and Sundaram \cite{34} proposed
a polynomial approximation of the input signal for accelerated approximate
Fast Fourier Transform (FFT) computations. Di Stefano and Mattoccia
\cite{118} presented an accelerated normalized spatial-domain cross-correlation
mechanism, with partial check according to an upper bound. Finally,
Kadyrov and Petrou \cite{82} and Anastasia and Andreopoulos \cite{81}
showed that it is possible to perform accelerated 2D convolution/cross-correlation
by piecewise packing of the input image data into a compact representation
when the algorithm utilizes integer inputs.

A third category of research advances on GEMM and CONV energy and
processing throughput adaptation is focusing on\emph{ specific error-tolerant
applications}, such as video codecs, image processing and signal processing
operations in custom hardware designs \cite{78,79,107,108,109,106,105,104,110,115}.
Beyond their reliance to specialized hardware or circuit design for
complexity--precision scalability of GEMM and CONV kernels, many such
approaches also tend to be \emph{algorithm-specific}. That is, they
use predetermined \textquotedblleft{}quality levels\textquotedblright{}
or \textquotedblleft{}profiles\textquotedblright{} of algorithmic
or system adjustment, e.g.: switching to simpler transforms or simplifying
algebraic operations \cite{15}\cite{120}\cite{121}, limiting the
operating precision of the algorithm implementation in a static manner
in order to satisfy hardware or processing constraints \cite{122},
or exploiting the structure of matrices in sparse matrix problems
\cite{123}. Previous research efforts by our group in image processing
systems \cite{30,81} were also algorithm-specific and, importantly,
no precision-controlled acceleration of linear operations was proposed.
For these reasons, many existing proposals of this category \cite{124,125,8,126,74,34}
are either based on complexity models or custom VLSI designs and cannot
be easily generalized to mainstream digital signal processors or high-performance
computing clusters. 

Overall, all current approaches for precision--energy--throughput
scaling of GEMM and CONV kernels appear to be limited by one or more
of the following: \emph{(i)} adaptation is only done at the process
level (e.g. results of entire tasks are dropped); \emph{(ii)} the
proposed methods are tailored to specific algorithms (e.g. image filtering
or specific signal transforms); \emph{(iii)} special-purpose hardware
is required and optimized deployment via mainstream processors with
streaming single-instruction multiple-data (SIMD) extensions is not
possible.

\subsection{Contribution }

This paper proposes an approach to scale precision, energy and throughput
(PET) scaling in GEMM and CONV kernels that form the dominant compute
and memory-intensive processing within broad classes of image/audio
recognition or matching systems. Our proposal is applicable to GEMM
and CONV kernels running on commercial off-the-shelf processors and,
via PET scaling, it is shown to significantly outperform state-of-the-art
deployments on such processors. Importantly, PET scaling in our approach
is done with straightforward selection of a few parameters that are
software-adjustable. Finally, our approach is not limited to a specific
algorithm or application; rather it is applicable to a large range
of ET applications based on GEMM and CONV kernels. 

\begin{figure}[tbh]
\begin{centering}
\includegraphics[scale=0.33]{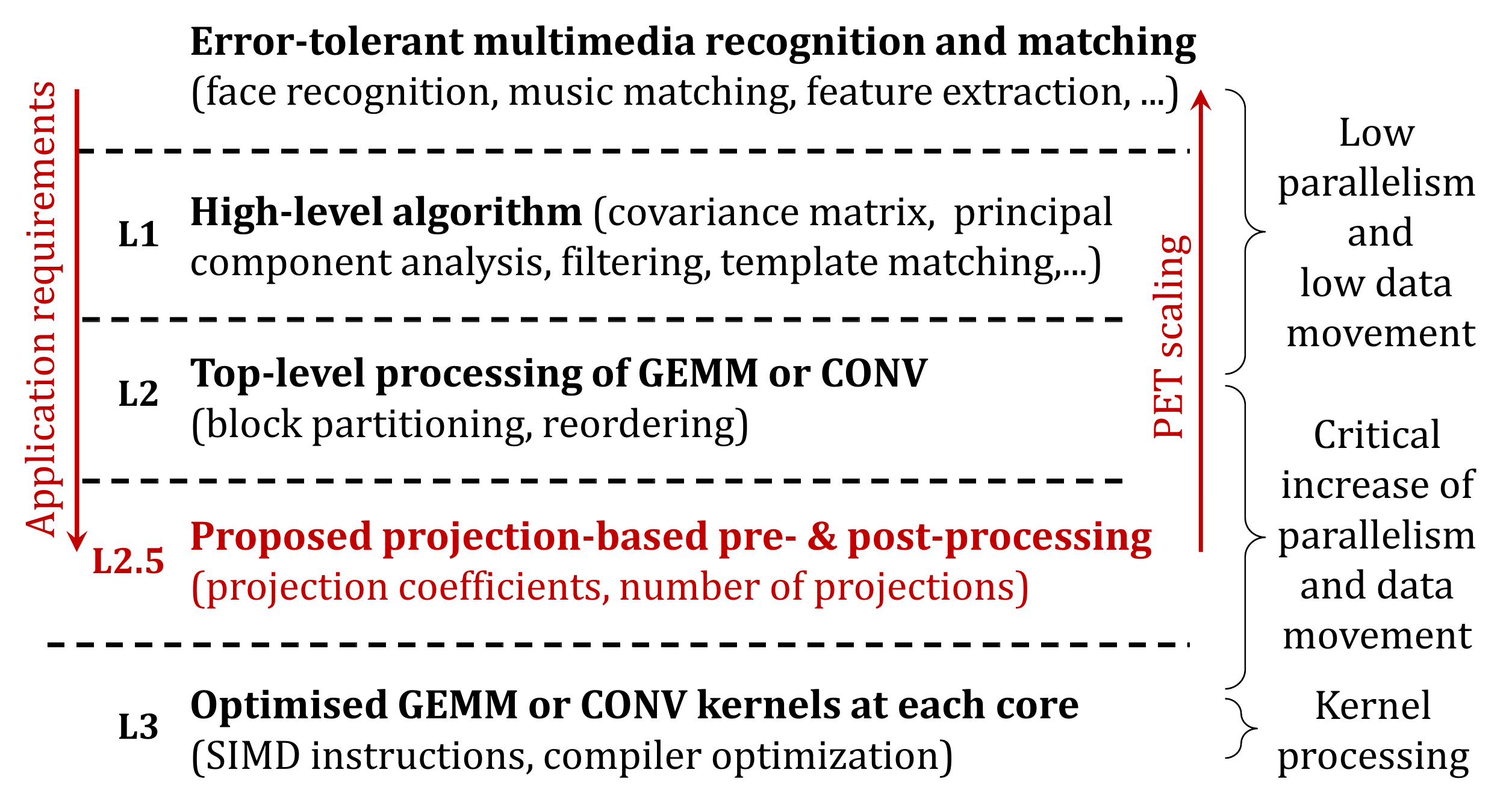}
\par\end{centering}

\caption{Proposed work positioned within the execution environment of ET multimedia
applications.\label{fig:framework}}
\end{figure}

To illustrate how these important advantages are achieved by our proposal,
Figure \ref{fig:framework} presents a schematic layering of the execution
of typical compute and memory-intensive ET multimedia applications
on high-performance and embedded systems. As shown in the figure,
between L2 and L3, a partitioning \cite{6} (or reordering \cite{68,69})
of the input data takes place and each data block is assigned to a
kernel-processing core (or thread) for memory-efficient (and, possibly,
concurrent) realization of subsets of GEMM and CONV computations.
Each core returns its output block of results to the top-level processing
of L2 and all blocks are assembled together to be returned to the
high-level algorithm. Parallelism and data movement to and from cores
tend to increase drastically between L2 and L3. 

When aiming for high-throughput/low-energy performance, the critical
issues of the execution environment of Figure \ref{fig:framework}
are \cite{111,68,69}: \emph{(i)} the data movement to/from cores;
\emph{(ii)} the processing time and energy consumption per core; \emph{(iii)}
the limited concurrency when the top-level processing allows for only
a few blocks. These issues are addressed in our proposal by viewing
the process between L2 and L3 as a \emph{computation channel }\cite{6}
that returns approximate results. All current approaches correspond
to the least-efficient, \textquotedblleft{}lossless\textquotedblright{},
mode (i.e. typically 32-bit floating-point accuracy), which will typically
be unable to accommodate timing and/or energy constraints imposed
by the application. It is proposed to create highly-efficient, \textquotedblleft{}lossy\textquotedblright{},
modes for pre- and post-processing of streams via projection techniques
(L2.5 of Figure \ref{fig:framework}). This is achieved by: \emph{(i)}
partitioning and reordering inputs in L2 to move them to each core
for kernel processing; \emph{(ii)} converting them into multiple,
compact, representations allowing for reduced data movement, increased
concurrency and fast recovery of approximate results from only a few
cores.

\subsection{Paper Organization }

In Section \ref{sec:2}, we review the top-level processing of GEMM
and CONV kernels considered in this paper. Section \ref{sec:3} presents
the proposed projections-based data compaction method within GEMM
and CONV kernels. Section \ref{sec:4} presents performance benchmarks
for the proposed method in terms of precision, energy consumption
and processing throughput attained on the recently-introduced ARM
Cortex A15 processor. In addition, comparisons against both the original
(i.e. non projections-based) kernels, as well as state-of-the-art
GEMM and CONV kernels from third parties, are carried out. Section
\ref{sec:5--} demonstrates the ability of the proposed approach to
achieve substantial resource--precision adaptation within two error-tolerant
multimedia recognition and matching applications. Finally, Section
\ref{sec:7} concludes the paper. 

\section{Overview of Top-level Processing of GEMM and CONV Kernels\label{sec:2}}

This section outlines the key aspects of data partitioning and reordering
within the GEMM and CONV kernels under consideration. Specifically,
the accessing and partitioning order of the input data streams is
shown, in order to provide the context under which the proposed projections-based
mechanisms are deployed. The nomenclature summary of this paper is
given in Table~\ref{tab:Nomenclature-table.1}.

\begin{table}[tbh]
\noindent \centering{}\caption{\label{tab:Nomenclature-table.1}Nomenclature table.}
\begin{tabular}{>{\centering}m{0.1\columnwidth}>{\raggedright}p{0.65\columnwidth}}
\multicolumn{1}{>{\centering}m{0.15\columnwidth}}{Symbol } & \multicolumn{1}{>{\raggedright}m{0.6\columnwidth}}{Definition}\tabularnewline
\midrule
\centering{}$M$, $K$, $W$ & Input matrix or signal size parameters\tabularnewline
\midrule
$N$ & Kernel size (e.g. $N\times N$ subblock in GEMM or $N$-sample convolution
kernel)\tabularnewline
\midrule
$L$ & Total number of projections used\tabularnewline
\midrule
$\mathbf{A}_{i}$, $\mathbf{a}$, $\mathbf{A^{\mathbf{T}}}$ & Boldface uppercase and lowercase letters indicate matrices and vectors,
respectively; superscript $\text{T}$ denotes transposition\tabularnewline
\midrule
$\widehat{r}_{\text{sub}}$ & Reconstruction of result $r$ using subset ``$\text{sub}$'' of
projections\tabularnewline
\midrule
$a\left[0,0\right]$ & Italicized lowercase letters indicate elements of corresponding matrices
or vectors (with the enumeration starting from zero)\tabularnewline
\midrule
$\left\Vert \mathbf{A}\right\Vert _{F}$ & Frobenius norm\tabularnewline
\midrule
\textsf{a$\leftarrow$instr(b,c)} & Indicates assignment of result to variable \textsf{a} after performing
instruction \textsf{instr} using \textsf{b} and \textsf{c} (the meaning
of the instruction is identifiable from the context) in pseudocode
listings\tabularnewline
\bottomrule
\end{tabular}
\end{table}

\subsection{Brief Review of Block Processing within GEMM }

\begin{figure}[tbh]
\begin{centering}
\includegraphics[scale=0.34]{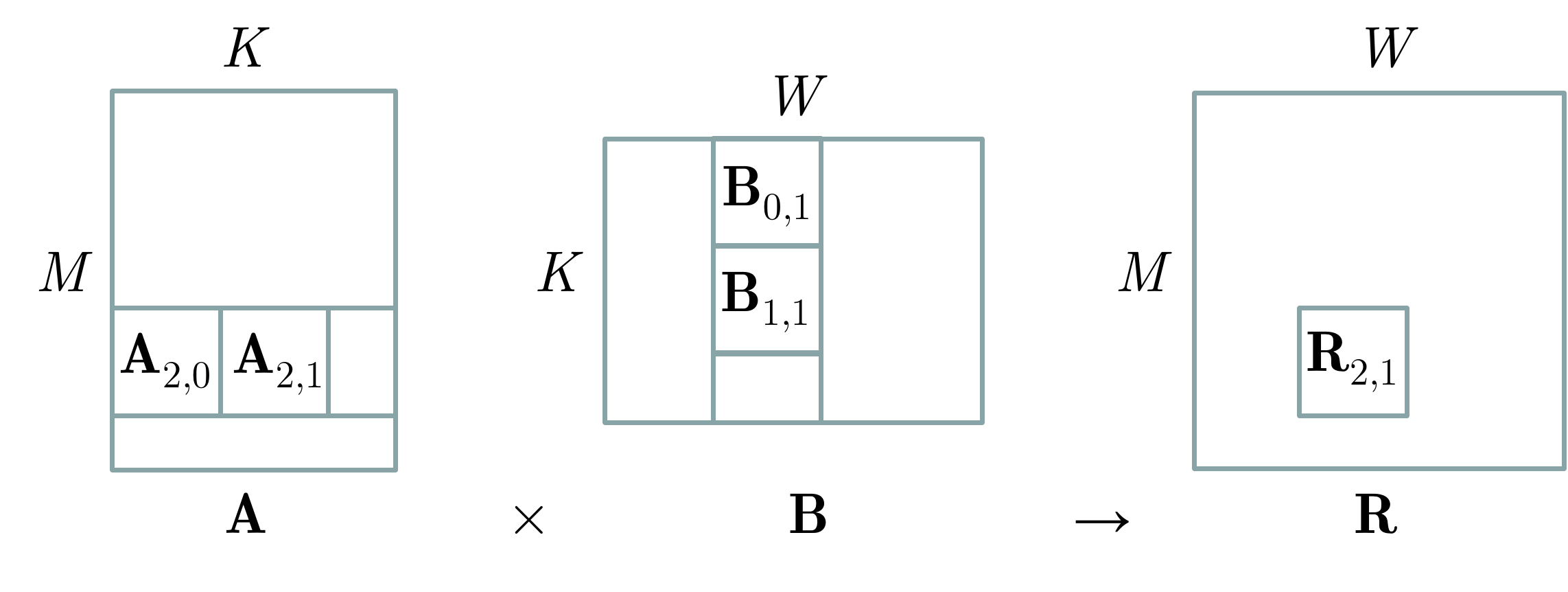} 
\par\end{centering}

\caption{Top-level processing of GEMM highlighting the input subblocks involved
in the example subblock result $\mathbf{R}_{2,1}$. \label{fig:GEMM operation}}
\end{figure}

Consider the standard GEMM design depicted in Figure \ref{fig:GEMM operation},
following the general flow found in optimized MKL designs \cite{68,69}.
The application invokes GEMM for an $M\times K$ by $K\times W$ matrix
multiplication that is further subdivided into $N\times N$ \textquotedblleft{}inner-kernel\textquotedblright{}
matrix products. For our approach, $N$ is specified by ($k\in\mathbb{N}^{\star}$):

\begin{equation}
N=2k\times\frac{\text{SIMD}_{\text{bits}}}{b_{\text{repr}}}\label{eq:N-definition}
\end{equation}
with: $\text{SIMD}_{\text{bits}}$ the number of bits of each SIMD
register ($\text{SIMD}_{\text{bits}}=128$ in this work); $b_{\text{repr}}=32$
the number of bits for floating-point or integer (fixed-point) representations.
The inner-kernel result, $\mathbf{R}_{2,1}$, of the example shown
in Figure \ref{fig:GEMM operation} comprises the sum of multiple
subblock multiplications $\mathbf{A}_{2,n}\mathbf{B}_{n,1}$, and
is given by: 

\begin{equation}
\mathbf{R}_{2,1}=\sum_{n=0}^{\frac{K}{N}-1}\mathbf{A}_{2,n}\mathbf{B}_{n,1}.\label{eq:matrix-products}
\end{equation}

If the matrices\textquoteright{} dimensions are not multiples of $N$,
some \textquotedblleft{}cleanup\textquotedblright{} code \cite{68,69}
is applied at the borders to complete the inner-kernel results of
the overall matrix multiplication. This separation into top-level
processing and subblock-level processing is done for efficient cache
utilization \cite{69,88}. Specifically, during the initial data access
of GEMM for top-level processing, data in matrix $\mathbf{A}$ and
$\mathbf{B}$ is reordered into block major format: for each $N\times N$
pair of subblocks $\mathbf{A}_{i,n}$ and $\mathbf{B}_{n,j}$ multiplied
to produce inner-kernel result $\mathbf{R}_{i,j}$, $0\leq n<\frac{K}{N}$,
$0\leq i<\frac{M}{N}$, $0\leq j<\frac{W}{N}$, the input data within
$\mathbf{A}_{i,n}$ and $\mathbf{B}_{n,j}$ is reordered in rowwise
and columnwise raster manner, respectively. Thus, sequential data
accesses are performed during each subblock matrix multiplication
and this enables the use of SIMD instructions, thereby leading to
significant acceleration. The appropriate value for the subblock dimension,
$N$, can be established for each architecture following an automated
process at compile time (e.g. via test runs {[}69{]}).

Our approach intercepts the subblock-based rowwise and columnwise
raster ordering (exploiting the fact that the input data subblock
is accessed anyway) in order to perform low-complexity linear projections
to the input rows and columns prior to the performance of individual
GEMMs within the projected data. In conjunction with the fact that
the proposed approach does not alter the top-level processing of the
standard GEMM, in the remainder of the paper we only refer to a single
subblock product. For notational simplicity, we remove the indices
from subblock product $\mathbf{A}_{i,n}\mathbf{B}_{n,j}$.

\subsection{Brief Review of Overlap-save Processing within CONV}

Consider the discrete convolution of two 1D signals, $\mathbf{s}_{\text{in}}$
and $\mathbf{k}$, producing the output signal, $\mathbf{r}_{\text{out}}$: 

\begin{equation}
\mathbf{r}_{\text{out}}=\mathbf{s}_{\text{in}}\star\mathbf{k}\Longleftrightarrow\forall m:\; r_{\text{out}}\left[m\right]=\sum_{n=0}^{N-1}s_{\text{in}}\left[n\right]k\left[m-n\right].\label{eq:convolution-definition}
\end{equation}
The signal with the smallest time-domain support is considered to
be the kernel, $\mathbf{k}$, and the other signal, $\mathbf{s}_{\text{in}}$,
is the input. Assuming $\mathbf{s}_{\text{in}}$ is periodic with
period $N$, circular convolution of period $N$ can be expressed
by:

\begin{eqnarray}
\mathbf{r}_{\text{out}} & = & \left(\mathbf{s}_{\text{in}}\circledast\mathbf{k}\right)_{N}\Longleftrightarrow\nonumber \\
\forall m:\; r_{\text{out}}\left[m\right] & = & \sum_{n=0}^{N-1}\left(\sum_{p=-\infty}^{\infty}s_{\text{in}}\left[n+pN\right]\right)k\left[m-n\right]\label{eq:circular-convolution-definition}
\end{eqnarray}
Finally, discrete cross-correlation and circular cross-correlation
can be obtained by replacing $k\left[m-n\right]$ with $k\left[m+n\right]$
in \eqref{eq:convolution-definition} and \eqref{eq:circular-convolution-definition}. 

\begin{figure}[tbh]
\begin{centering}
\includegraphics[scale=0.41]{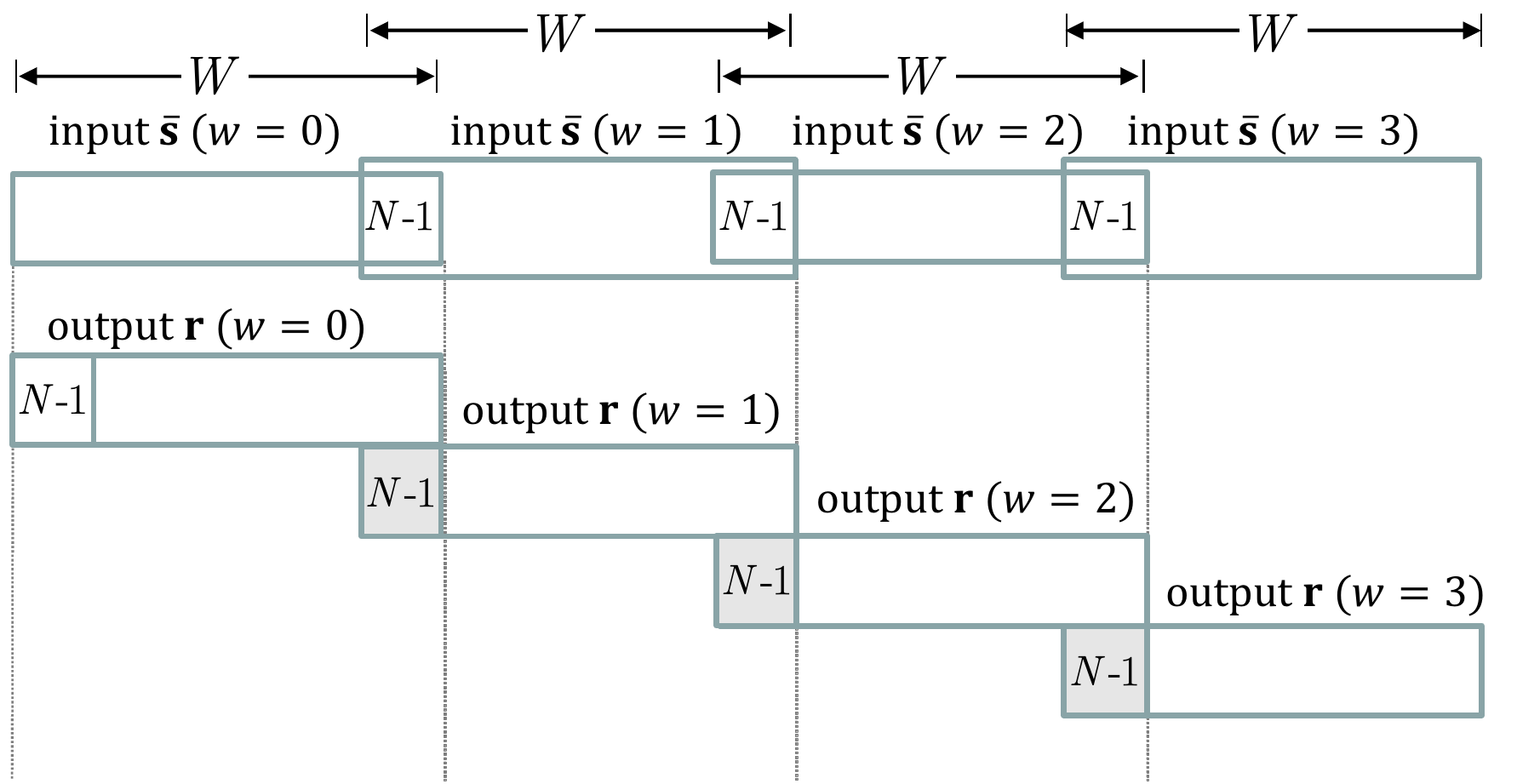} 
\par\end{centering}

\caption{Top-level processing of CONV highlighting the overlap-save method.
\label{fig:CONV operation}}
\end{figure}

As shown in Figure \ref{fig:CONV operation}, practical implementations
of convolution of a long input signal with an $N$-sample kernel $\mathbf{k}$
will subdivide the input into several partially-overlapping blocks---of
$W$ samples each (vector $\mathbf{s}$)---prior to the actual convolution.
Each individual signal block $\mathbf{s}$ is independently convolved
with the kernel and the resulting blocks ($\mathbf{r}$) are assembled
together to give the result of the convolution. This is the well-known
overlap-save method \cite{112}, performed for efficient cache utilization
and increased concurrency, with the degree of concurrency and the
processing delay depending on $W$. The optimal value of $W$ for
the utilized architecture can be derived based on offline experimentation,
e.g., during the routine compilation or via offline experiments with
the target processor \cite{7}. 

Our approach exploits the fact that overlap-save CONV accesses blocks
of data (in order to subdivide the input) and applies low-complexity
projection operations during this process. Similarly, as for the case
of GEMM, in the next section we shall only be presenting the proposed
method for one block of $W$ samples and, for notational simplicity,
we shall not retain the block index but rather consider it to be the
entire input signal $\mathbf{s}_{\text{in}}$.

\section{PET Scaling of Numerical Kernels via Linear Projections \label{sec:3}}

We first present an example of how projections in numerical kernels
produce a hierarchical representation of inner-product computations,
which comprise the core operation within both kernels. In Subsection
\ref{sub:3.2} we elaborate on the deployment of the proposed projections-based
resource--precision scaling within high-performance GEMM and CONV
kernels. Finally, in Subsection \ref{sub:3.3} we quantify its multiply--accumulate
(MAC) operations and required data transfers between top-level and
kernel processing against the standard kernel realization that does
not use projections.

\subsection{Illustration of the Basic Concept\label{sub:3.1}}

\begin{figure}[tbh]
\begin{centering}
\includegraphics[scale=0.34]{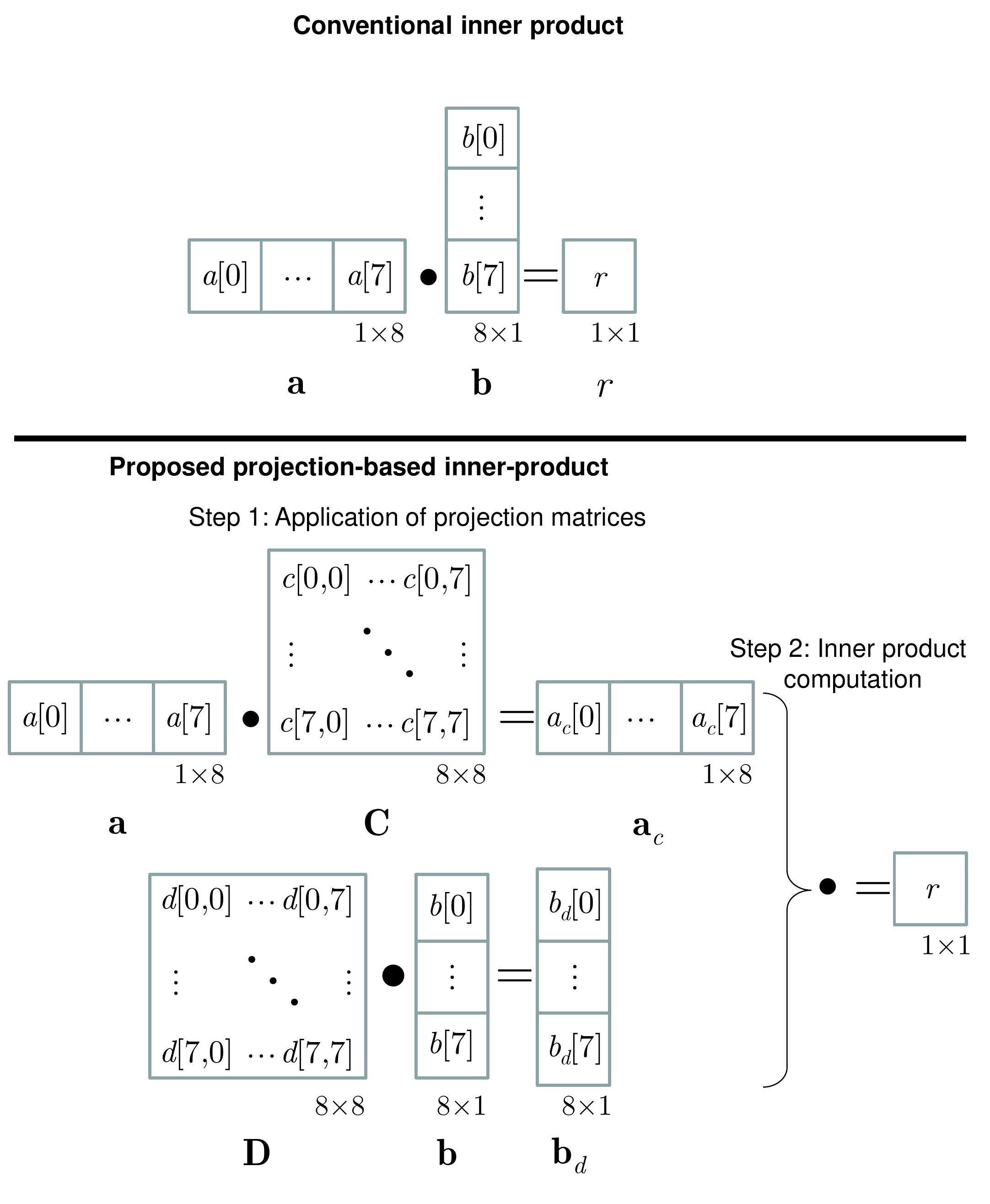} 
\par\end{centering}

\caption{Eight-sample inner product computation using $8\times8$ projection
matrices \textbf{$\mathbf{C}$} and $\mathbf{D}$. Top half: inner
product computation. Bottom half: projections-based inner product.
\label{fig:conv-inner-product}}
\end{figure}

Consider the calculation of an inner product $r=\mathbf{a}\mathbf{b}$,
such as the one illustrated in the example of the top half of Figure
\ref{fig:conv-inner-product}. We can apply projection matrices $\mathbf{C}$
and $\mathbf{D}$ to the inputs by:

\begin{equation}
\mathbf{a}_{c}=\mathbf{a}\mathbf{C},\;\;\mathbf{b}_{d}=\mathbf{D}\mathbf{b}\label{eq:basic-linear-projections}
\end{equation}

If $\mathbf{C}$ is an invertible square matrix and we set $\mathbf{D}=\mathbf{C}^{-1}$,
the inner product can take place using the projected vectors, since:

\begin{equation}
\mathbf{a}_{c}\mathbf{b}_{d}=\left(\mathbf{a}\mathbf{C}\right)\left(\mathbf{D}\mathbf{b}\right)=\mathbf{a}\mathbf{\left(CD\right)}\mathbf{b}=\mathbf{ab}=r\label{eq:inner-product-projected}
\end{equation}
which is illustrated in the bottom half of Figure \ref{fig:conv-inner-product}.
If one ignores the cost of performing the projections of \eqref{eq:basic-linear-projections},
the inner product of \eqref{eq:inner-product-projected} incurs the
same computational effort as the original inner product%
\footnote{The implementation cost of \eqref{eq:basic-linear-projections} is
certainly non-negligible in this example. However, in the next subsection
we illustrate that one can find an appropriate balance between the
number of projections performed and the subblock size in GEMM, or
kernel size in CONV, $N$, in order for this cost to be reasonably
small. %
}. Importantly, the projection matrices can prioritize the computation
of the result since, if appropriately selected, they can concentrate
the energy of the inputs in the first few elements. For example, considering
that the input vectors $\mathbf{a}$ and $\mathbf{b}$ comprise image
or signal data with energy concentrated in low frequencies and \textbf{$\mathbf{C}$}
is chosen as the $L$-point discrete cosine transform (DCT) transform
($L=8$, $\forall i,j:\:0\leq i,j<8$):

\begin{equation}
c\left[i,j\right]=\cos\left[\frac{\pi}{L}\left(i+\frac{1}{2}\right)j\right],\label{eq:DCT definition}
\end{equation}
if we only perform $\widehat{r}_{\text{DC}}=a_{c}\left[0\right]b_{d}\left[0\right]$,
this corresponds to reconstructing the ``DC component'' of the entire
inner product of \eqref{eq:inner-product-projected}. In addition,
this can optionally be incremented up to the eighth harmonic (i.e.,
reconstructing $r$ up to---and including---the eighth harmonic) by:

\begin{equation}
\widehat{r}_{\text{full}}=\sum_{l=0}^{7}a_{c}\left[l\right]b_{d}\left[l\right].\label{eq:increment-projections}
\end{equation}
with $\widehat{r}_{\text{full}}=r$ barring numerical approximation
error. The computation of each harmonic $a_{c}\left[l\right]b_{d}\left[l\right]$
can be assigned to a different processor and the accumulation of \eqref{eq:increment-projections}
is optional: if less than all eight harmonics are accumulated, the
precision of the result is expected to degrade gracefully. This reduces
the energy consumption and data transfers to and from processors,
or increases processing throughput if a single processor is used.

For a population of $N$ results ($0\leq n<N$), e.g., $r\left[n\right]=\mathbf{a}\mathbf{b}\left[n\right]$,
the signal-to-noise ratio (SNR) between $r\left[0\right],\,\ldots,\, r\left[N-1\right]$
and $\widehat{r}_{\textrm{sub}}\left[0\right],\,\ldots,\,\widehat{r}_{\textrm{sub}}\left[N-1\right]$
(results computed in single-precision floating point vs. results reconstructed
from subset ``$\text{sub}$'' of projections), is given by: 

\begin{equation}
\text{SNR}=10\log_{10}\left(\frac{\sum_{n=0}^{N-1}r^{2}\left[n\right]}{\sum_{n=0}^{N-1}\left(r\left[n\right]-\widehat{r}_{\textrm{sub}}\left[n\right]\right)^{2}}\right).\label{eq:SNR}
\end{equation}
If the projection matrices concentrate the energy of the input data
in these projections, then the SNR of \eqref{eq:SNR} can be adequately
high for an error-tolerant multimedia application. 

The extension of this simple example to inner products performed within
matrix product computations is relatively straightforward to envisage.
However, in the case of convolution/cross-correlation, due to the
translations performed during the calculation of the results, we first
need to define the cyclic permutation matrix comprising $N\times N$
elements \cite{112}:\\
\begin{equation}
\mathbf{P}_{n}=\begin{bmatrix}\mathbf{0} & \mathbf{I}_{n}\\
\mathbf{I}_{N-n} & \mathbf{0}
\end{bmatrix}
\end{equation}
with: $0\leq n<N$, $\mathbf{I}_{k}$ the $k\times k$ identity matrix
and $\mathbf{0}$ the zero matrix whose dimensions are identifiable
from the context.

We can then define the projections-based circular cross-correlation
operation for the example of Figure \ref{fig:conv-inner-product}
based on the following steps: 
\begin{enumerate}
\item For all translations $n$, $0\leq n<8$, derive the translated-and-projected
inputs:\\
\begin{equation}
\mathbf{a}_{c}\left[n\right]=\mathbf{a}\mathbf{P}_{n}\mathbf{C},\label{eq:all-translations-of-one-input}
\end{equation}
with $\mathbf{a}_{c}\left[n\right]$ the $1\times8$ vector corresponding
to the projection of the $n$th cyclic translation (permutation) of
$\mathbf{a}$.
\item Derive the projected input $\mathbf{b}_{d}$ by:\\
\begin{equation}
\mathbf{b}_{d}=\mathbf{Db}.\label{eq:all-samples-of-other-projected-input}
\end{equation}

\item Reconstruct the $\left(7-n\right)$th sample of the output by ($0\leq n<8$):\\
 
\begin{equation}
\widehat{r}_{\text{full}}\left[7-n\right]=\mathbf{a}_{c}\left[n\right]\mathbf{b}_{d}=\sum_{l=0}^{7}a_{c}\left[n,l\right]b_{d}\left[l\right].\label{eq:all-results-with-all-projections}
\end{equation}

\end{enumerate}
Circular convolution can be defined following the same steps if we
reverse the order of either $\mathbf{a}$ or $\mathbf{b}$. Moreover,
discrete convolution and cross-correlation are defined by these steps
if extension with zeros is performed in \eqref{eq:all-translations-of-one-input}
instead of cyclic permutations. Notice that well-known acceleration
techniques like the FFT can be applied in \eqref{eq:all-results-with-all-projections}
since, when considering all translations $n$, \eqref{eq:all-results-with-all-projections}
comprises a variant of cross-correlation. 

In the case of convolution, we have two options to scale performance.
Firstly, we can opt to omit the calculation of some of the results
of \eqref{eq:all-results-with-all-projections} and instead interpolate
them from neighboring results, e.g. compute every other result and
replace the missing ones by averaging the neighboring results. Secondly,
we can opt to omit the calculation of some of the higher-numbered
products within the summation of \eqref{eq:all-results-with-all-projections},
which correspond to the higher harmonics of the translated-and-projected
inputs. Both options will lead to approximate results, with the resulting
error being quantified by the SNR calculation of \eqref{eq:SNR}.
For instance, for the case of $\mathbf{C}$ being the $8\times8$
DCT transform, we can reconstruct the DC component of the $\left(7-n\right)$th
sample of the output by ($0\leq n<8$):

\begin{equation}
\widehat{r}_{\text{DC}}\left[7-n\right]=a_{c}\left[n,0\right]b_{d}\left[0\right].\label{eq:all-results-with-DC-projections}
\end{equation}

\subsection{Application of the Concept within the Top-level Data Partitioning
and Reordering of GEMM and CONV\label{sub:3.2}}

For efficient deployment of projections-based processing within the
blocked GEMM or CONV kernels, we must: \emph{(i)} align the projection
matrix size to the block size of each kernel and \emph{(ii)} ensure
the entire process is performed without breaking the access pattern
of the data blocking (and possibly reordering) of the top-level processing
of each kernel. The latter is important because this means the entire
reordering and projections approach can be performed in a streaming
manner, i.e. with high-performance SIMD instructions\emph{.} 

Assuming that the inner-kernel size comprises $N$ samples and the
projection matrix comprises $L\times L$ coefficients, the first condition
is satisfied if $N$ is divisable by $L$. For example, the values
used in our experiments are: $N\in\left\{ 144,600,1200\right\} $
and $L\in\left\{ 2,8\right\} $. 

Concerning the second condition, we first define the mathematical
process of consecutive application of the projection kernels within
each $N\times N$ GEMM subblock or each $N$-sample convolution kernel.
This is achieved by defining the $N\times N$ block-diagonal matrices:

\begin{equation}
\mathbf{C}_{N}=\begin{bmatrix}\mathbf{C} & \mathbf{0} & \cdots & \mathbf{0}\\
\mathbf{0} & \mathbf{C} &  & \mathbf{0}\\
\vdots &  & \ddots & \vdots\\
\mathbf{0} & \mathbf{0} & \cdots & \mathbf{C}
\end{bmatrix},\;\mathbf{D}_{N}=\begin{bmatrix}\mathbf{D} & \mathbf{0} & \cdots & \mathbf{0}\\
\mathbf{0} & \mathbf{D} &  & \mathbf{0}\\
\vdots &  & \ddots & \vdots\\
\mathbf{0} & \mathbf{0} & \cdots & \mathbf{D}
\end{bmatrix},
\end{equation}
with: $\mathbf{C}$ and $\mathbf{D}$ the $L\times L$ projection
matrices, $\mathbf{C}$ comprising any invertible matrix and $\mathbf{D}=\mathbf{C}^{-1}$.
The mathematical application of the projection process then follows
the exposition of the previous subsection, albeit ignoring all elements
of the projection operations that contain zero coefficients.

\begin{figure*}[tbh]
\begin{centering}
\includegraphics[scale=0.34]{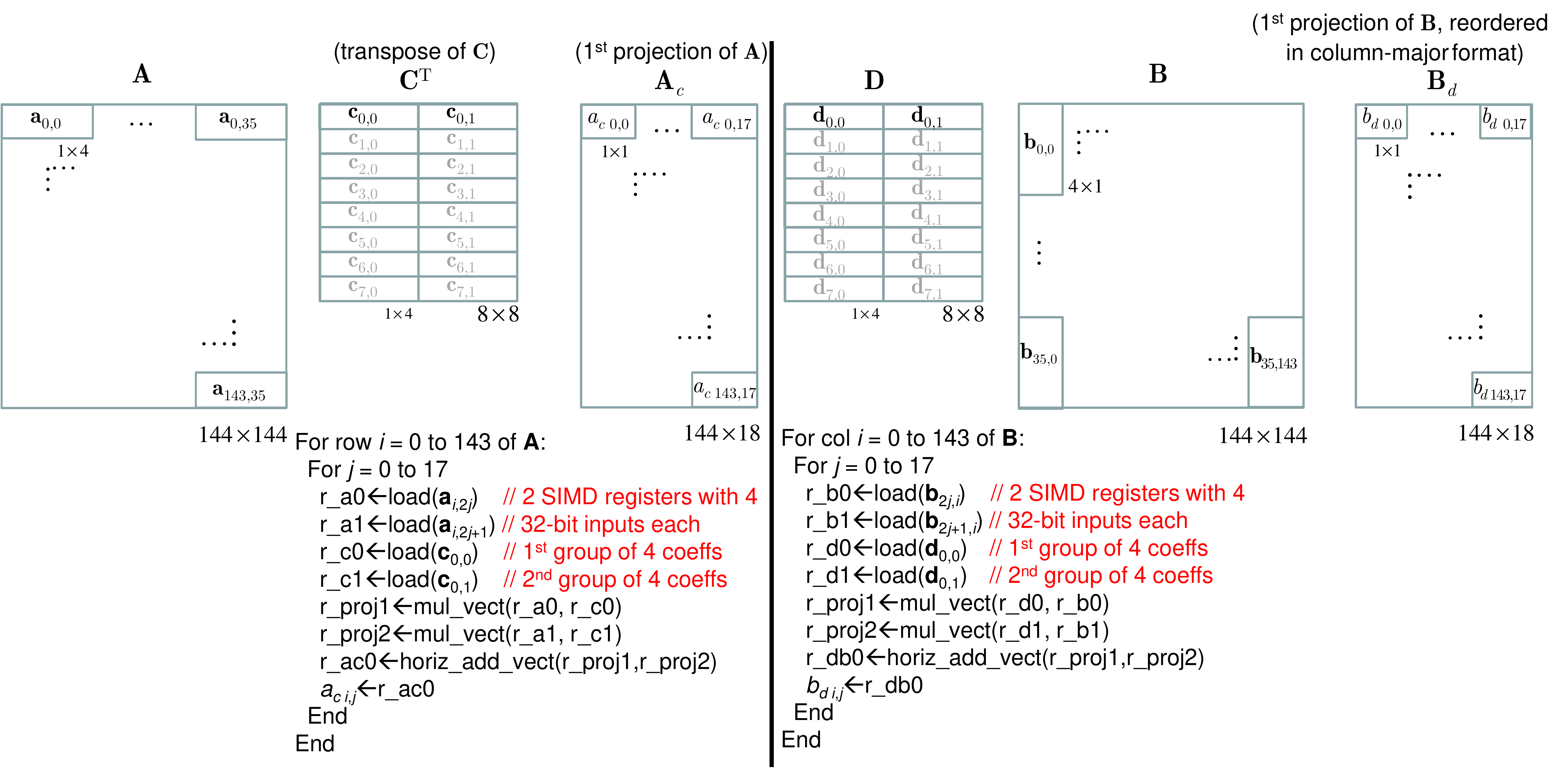} 
\par\end{centering}

\caption{Application of first projection of $8\times8$ projection matrices
\textbf{$\mathbf{C}$} and $\mathbf{D}$ during block-major reordering
in GEMM under SIMD registers storing four 32-bit elements. Left half:
right multiplication of blocks within the rows of $\mathbf{A}$ by
all elements of the first row of $\mathbf{C}^{\text{T}}$ (which is
equivalent to left-multiplying with the first column of $\mathbf{C}$).
Right half: left multiplication of blocks within the columns of $\mathbf{B}$
by the first row of $\mathbf{D}$. The grayed-out rows of $\mathbf{C}^{\text{T}}$
and $\mathbf{D}$ correspond to the subsequent projections. \label{fig:GEMM projection}}
\end{figure*}

In order to illustrate how this can be done following the access pattern
of the input data partitioning and reordering (and, more specifically,
using with SIMD instructions), Figure \ref{fig:GEMM projection} demonstrates
one projection operation during the block-major reordering performed
in GEMM. The figure illustrates the application of the first projection
vector (first row of $\mathbf{C}^{\text{T}}$ and first row of $\mathbf{D}$)
within a pair of subblocks of size $N\times N$. In this example,
we selected $N=144$ and $L=8$ (i.e. eight projections, comprising
eight coefficients each), which are the values used in our experiments.
In addition, we left-multiply each row of $\mathbf{A}$ with each
row of $\mathbf{C}^{\text{T}}$, which is equivalent to right-multiplying
the rows of $\mathbf{A}$ with the columns of $\mathbf{C}$ but it
is more efficient as all input elements are contiguous in memory (thereby
allowing for the use of SIMD instructions). This is illustrated in
the pseudocode of Figure \ref{fig:GEMM projection}, where we present
simplified SIMD instructions used in the inner loop of the projection
operation performed: \textsf{mult\_vect(r1,r2)} multiplies two SIMD
registers \textsf{r1} and \textsf{r2} and \textsf{horiz\_add\_vect(r1,r2)}
adds all eight elements within \textsf{r1} and \textsf{r2} (each SIMD
register%
\footnote{known as ``q-registers'' in the ARM Neon architecture%
} has four 32-bit elements). Specifically, the realization of the first
projection of the $j$th group of eight values in the $i$th row of
$\mathbf{A}$ is performed via the following (pseudocode of the left
part of Figure \ref{fig:GEMM projection}): 
\begin{itemize}
\item the first two load instructions of the inner \textsf{For} loop load
two pairs of four consecutive values of $\mathbf{A}$ into registers
\textsf{r\_a0} and \textsf{r\_a1}; 
\item the next two instructions load the two pairs of four consecutive projection
coefficients of the first row of $\mathbf{C}^{\text{T}}$ into registers
\textsf{r\_c0} and \textsf{r\_c1}; 
\item two vector multiplications are then carried out (\textsf{r\_a0} $\times$
\textsf{r\_c0} and \textsf{r\_a1} $\times$ \textsf{r\_c1}) and the
results are stored in registers \textsf{r\_proj1} and \textsf{r\_proj2}; 
\item the contents of these two registers are all added together to create
the $\left(i,j\right)$th element of $\mathbf{A}_{c}$.
\end{itemize}
The equivalent process is carried out for the realization of the first
projection of the $j$th group of eight values in the $i$th column
of $\mathbf{B}$ (shown in the pseudocode of the right part of Figure
\ref{fig:GEMM projection}). 

As shown in Figure \ref{fig:GEMM projection}, this process results
in a smaller GEMM product of dimensions $144\times18$ by $18\times144$.
All eight projections can be derived by using the subsequent (grayed-out)
rows of $\mathbf{C}^{\text{T}}$ and $\mathbf{D}$ and they can be
performed independently in eight different processing cores. This
results in: \emph{(i)} eight-fold increase of concurrency/data-level
parallelism within each subblock product, \emph{(ii)} reduced data
transfers to each core. Moreover, by computing only a small number
of projections, e.g. just one to three, this approach allows for graceful
degradation of the SNR of \eqref{eq:SNR} under energy and throughput
scaling. 

\begin{figure}[tbh]
\begin{centering}
\includegraphics[scale=0.34]{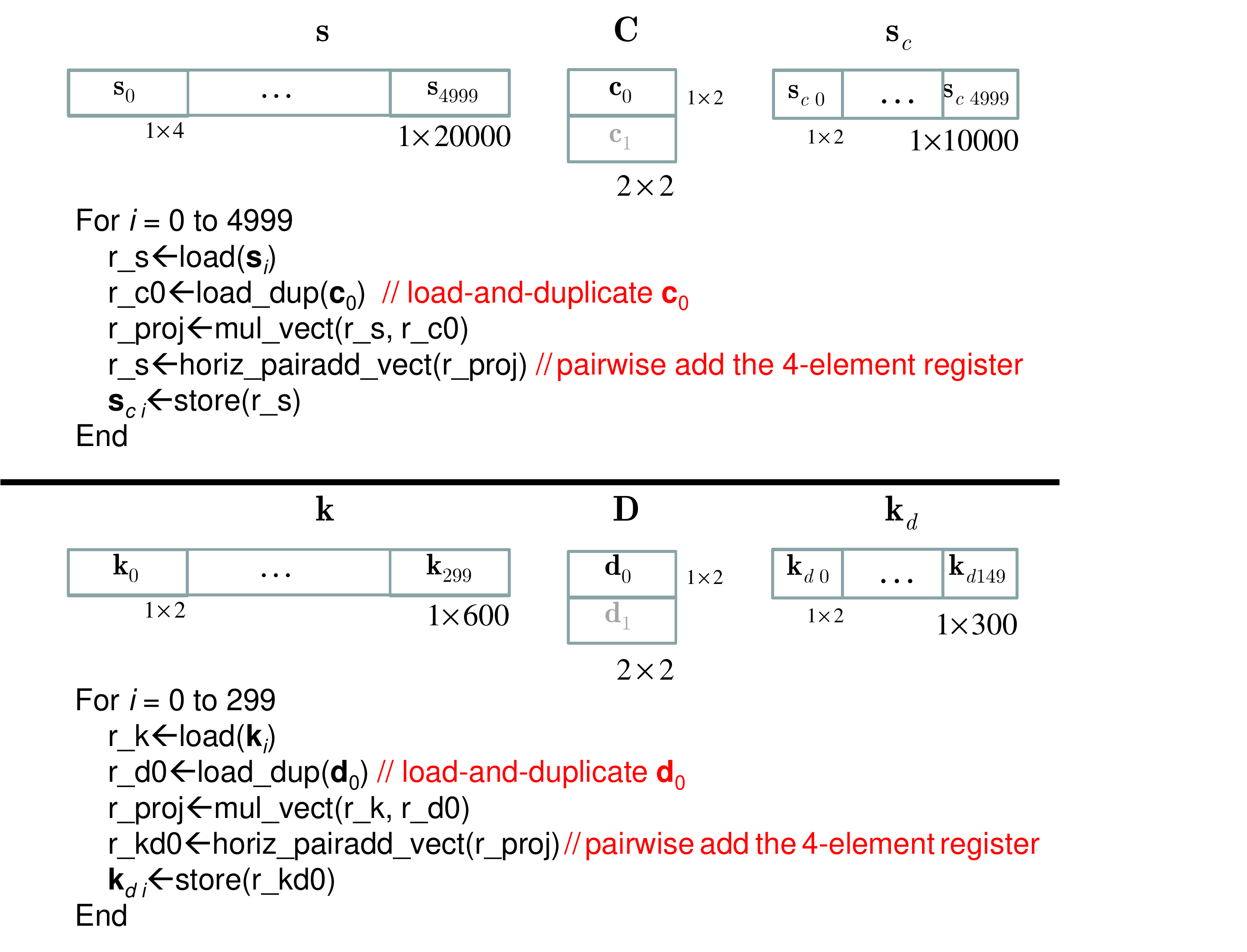} 
\par\end{centering}

\caption{Application of projection operations during blocking of CONV. Top
half: projection of input signal $\mathbf{s}$. Bottom half: projection
of kernel $\mathbf{k}$. \label{fig:CONV projection}}
\end{figure}

Concerning the signal block partitioning during the top-level processing
of CONV, in Figure \ref{fig:CONV projection} we demonstrate a single
projection operation applied to the input signal $\mathbf{s}$ and
kernel $\mathbf{k}$. Here, we utilize the following sizes for the
signal, kernel and projections: $W=20000$, $N=600$ and $L=2$, which
correspond to the values used in our experiments. In the pseudocode
of Figure \ref{fig:CONV projection}, \textsf{load\_dup(}$\mathbf{c}_{0}$\textsf{)}
loads the two elements of $\mathbf{c}_{0}$ and duplicates them within
one SIMD register and \textsf{horiz\_pairadd\_vect(r)} performs two
pairwise additions within the four elements of \textsf{r}. Specifically,
the realization of the projection of the $i$th group of four values
in $\mathbf{s}$ is performed via the following:
\begin{itemize}
\item the first load instruction of the \textsf{For} loop loads four consecutive
values of $\mathbf{s}$ into register \textsf{r\_s};
\item the second load instruction loads and duplicates the two values of
the first row of $\mathbf{C}$ into register\textsf{ r\_c0}; 
\item a vector multiplication is then carried out (\textsf{r\_s} $\times$
\textsf{r\_c0}) and the results are stored in register \textsf{r\_proj}; 
\item the contents of this register are all added together to create the
$i$th element of $\mathbf{s}_{c}$.
\end{itemize}
The equivalent process is carried out for the realization of the first
projection of the $i$th group of four values in kernel $\mathbf{k}$. 

Figure \ref{fig:GEMM projection} shows that the projection leads
to a convolution operation with half the number of samples in both
the input signal and the kernel ($\mathbf{s}_{c}$ and $\mathbf{k}_{d}$,
respectively), thereby asymptotically decreasing the arithmetic complexity
by a factor of four. This can be extended to higher gains if higher
values of $L$ are used.

\subsection{Computational and Memory Aspects of Projections-based GEMM and CONV\label{sub:3.3}}

The conventional (or ``plain'') GEMM kernel (i.e., without projections)
requires 
\begin{equation}
\text{C}_{\text{GEMM,plain}}\left(N\right)=N^{3}\label{eq:MAC_GEMM-plain}
\end{equation}
MAC operations for each pair of $N\times N$ subblocks $\mathbf{A}$
and $\mathbf{B}$. On the other hand, for deriving $\left(l+1\right)$
projections out of the $L$ possible ones ($0\leq l<L$), $2\left(l+1\right)N^{2}$
MAC operations are performed in the input subblocks, followed by $\frac{l+1}{L}N^{3}$
MAC operations for the $\left(l+1\right)$ smaller matrix products,
$\mathbf{A}_{c}\mathbf{B}_{d}$ (shown in Figure \ref{fig:GEMM projection}
for the first projection), and $lN^{2}$ accumulation operations to
produce the final results. Thus, in total,

\begin{equation}
\text{C}_{\text{GEMM,proj}}\left(N,l,L\right)=N^{2}\left[\frac{l+1}{L}N+3l+2\right]
\end{equation}
MAC operations are required for the proposed approach when performing
$\left(l+1\right)$ projections of $L$ coefficients each. In terms
of data transfer and storage requirements, the conventional GEMM requires
$\text{M}_{\text{GEMM,plain}}=2N^{2}b_{\text{repr}}$ bits to be transferred
to each GEMM subblock kernel {[}with $b_{\text{repr}}$ the number
of bits of the utilized numerical representation, defined as for \eqref{eq:N-definition}{]}
and the proposed approach requires $\text{M}_{\text{GEMM,proj}}=2\frac{l+1}{L}N^{2}b_{\text{repr}}$
bits to be transferred to the $\left(l+1\right)$ GEMM subblock kernels.
Thus, if $l<L-1$, the proposed approach reduces the memory transfer
and storage requirements by $\left(1-\frac{l+1}{L}\right)\times100\%$.

Concerning the CONV kernel, under the assumption of minimum-size signal
blocking for overlap-save operation \cite{7} (larger input signal
block sizes will have proportionally-higher requirements for all methods),
i.e. $W=3N+1$, the conventional CONV kernel (i.e. without projections)
requires \cite{7}: 
\begin{equation}
\text{C}_{\text{CONV,plain,time}}\left(N\right)=2N^{2}
\end{equation}
MAC operations for time-domain convolution/cross-correlation realization
and, approximately: 
\begin{equation}
\text{C}_{\text{CONV,plain,freq}}\left(N\right)=\left(45N+15\right)\log_{2}\left(3N+1\right)+3N+1\label{eq:MAC_CONV-plain-freq}
\end{equation}
MAC operations under a frequency-domain (FFT-based) realization. The
approximation of \eqref{eq:MAC_CONV-plain-freq} stems from the FFT
approximation formula of Franchetti \emph{et al} \cite{116}. Concerning
the proposed approach, the application of $\left(l+1\right)$ projections
($0\leq l<L$, each projection comprising $L$ coefficients) to both
the signal and kernel requires $\left(l+1\right)\left(4N+1\right)$
MAC operations, followed by $\left(l+1\right)$ CONV kernels applied
to the downsampled signals. Thus, the overall number of MAC operations
for time-domain and frequency-domain processing under the proposed
approach is: 
\begin{equation}
\text{C}_{\text{CONV,proj,time}}\left(N,l,L\right)=\left(l+1\right)\left(4N+1\right)+2\left(l+1\right)\left\lceil \frac{N}{L}\right\rceil ^{2}
\end{equation}
 and 
\begin{eqnarray}
\text{C}_{\text{CONV,proj,freq}}\left(N,l,L\right) & = & \left(l+1\right)\left(4N+1\right)+\left(l+1\right)\nonumber \\
 & \cdot & \left[\left(45\left\lceil \frac{N}{L}\right\rceil +15\right)\right.\label{eq:MAC_CONV-proj_freq}\\
 & \cdot & \left.\log_{2}\left(3\left\lceil \frac{N}{L}\right\rceil +1\right)+3\left\lceil \frac{N}{L}\right\rceil +1\right].\nonumber 
\end{eqnarray}

Finally, in terms of data transfer and storage requirements, the conventional
CONV kernel requires $\text{M}_{\text{CONV,plain}}=\left(4N+1\right)b_{\text{repr}}$
bits to be transferred to the CONV kernel, while the proposed approach
requires $\text{M}_{\text{CONV,proj}}=\left\lceil \frac{l+1}{L}\left(4N+1\right)\right\rceil b_{\text{repr}}$
bits to be transferred to the $\left(l+1\right)$ CONV kernels, thereby
leading to a reduction by $\left(1-\frac{l+1}{L}\right)\times100\%$
if $l<L-1$. 

\begin{figure}[tbh]
\begin{centering}
\includegraphics[scale=0.17]{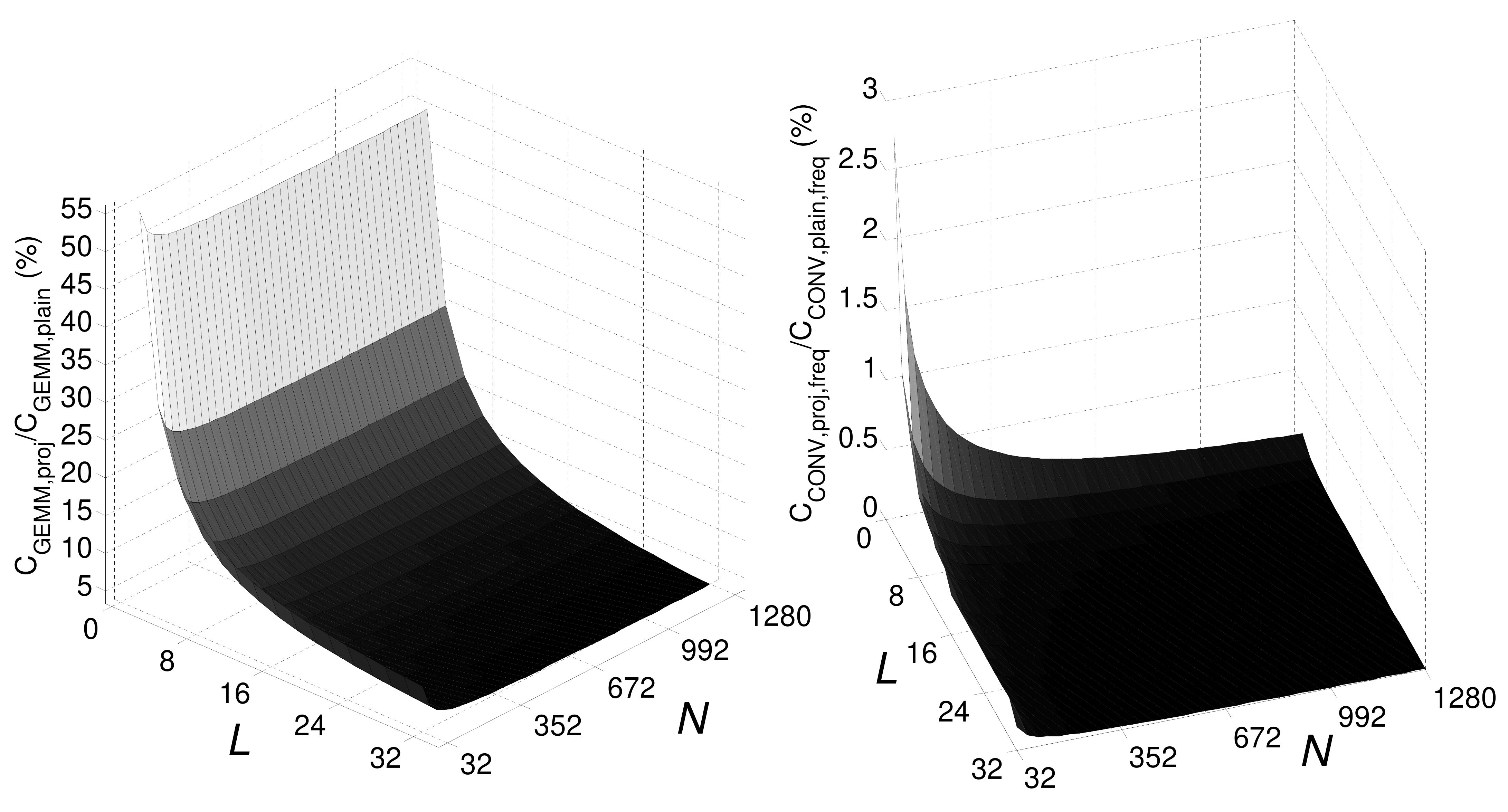} 
\par\end{centering}

\caption{Arithmetic complexity percentile ratios (proposed versus conventional)
for GEMM and CONV (frequency-domain) kernels of size $N$ using projection
size $L$ and performing a single projection. \label{fig:complexity}}
\end{figure}

Based on \eqref{eq:MAC_GEMM-plain}--\eqref{eq:MAC_CONV-proj_freq},
Figure \ref{fig:complexity} presents the ratios $\frac{\text{C}_{\text{GEMM,proj}}}{\text{C}_{\text{GEMM,plain}}}\times100\%$
and $\frac{\text{C}_{\text{CONV,proj,freq}}}{\text{C}_{\text{CONV,plain,freq}}}\times100\%$
for various values of $N$ and $L$ when performing only one projection
($l=0$). Evidently, the proposed approach is expected to lead to
substantial savings in arithmetic complexity, which in turn will translate
to increased throughput and energy efficiency in a real deployment.
This is experimentally verified in the next two sections, in conjunction
with the obtained precision within error-tolerant applications.

\section{Experimental Results\label{sec:4}}

We present results using the dual-core ARM Cortex A15 out-of-order
superscalar processor (ARM v7 instruction set, bare metal, only one
core was used and the other was powered down) with 32 KB L1 cache
(for instructions and data) and 4 MB L2 cache. This processor has
recently been integrated in popular System-on-Chip products marketed
for multimedia applications in smartphone and home entertainment environments,
such as Samsung's Exynox 5, Exynox 5 Octa, Apple TV and the Google
Chromebook portable computer. Similar results using our method were
also obtained in an Intel Core i7-4700MQ 2.40GHz processor, but are
omitted for brevity of exposition and also because energy consumption
could not be measured as accurately as for the ARM processor based
on the hardware available to us at the time of this writing.

The GEMM and CONV kernels were deployed on the ARM Cortex A15 using
C code with 32-bit floating-point Neon instructions (ARM SIMD extensions
utilizing the q registers of the processor) for accelerated processing.
All codes were compiled by the ARM Development Studio 5 (DS5) C compiler
under full optimization. Results were obtained using the ARM Versatile
Express board with the V2P-CA15 (ASIC A15 chip) daughter-board and
the ARM RealView ICE debugger unit. Dynamic and static power consumption
was measured directly in hardware by the ARM energy probe%
\footnote{For further information on the utilized tools, please see: \href{http://goo.gl/FVwrg}{http://goo.gl/FVwrg}
(ARM versatile express); \href{http://goo.gl/M3Crk}{http://goo.gl/M3Crk}
(ARM Neon architecture); \href{http://www.arm.com/products/tools/software-tools/ds-5/index.php}{http://www.arm.com/products/tools/software-tools/ds-5/index.php}
(ARM Development Studio 5); \href{http://goo.gl/YXYFB}{http://goo.gl/YXYFB}
(ARM Energy Probe).%
}. The board allows for dynamic voltage/frequency scaling (DVFS) between
$V_{\text{dd}}=0.6$ V at $0.6$ GHz and $V_{\text{dd}}=0.85$ V at
$1.2$ GHz at room temperature. In order to increase the reliability
of our results, each experiment was performed 100 times using representative
input data from image and audio streams normalized between $\left[-1,1\right]$;
the presented precision--energy--throughput results stem from averages
over all runs. Precision is measured in terms of SNR (dB) against
the result computed by the conventional GEMM and CONV kernels in single-precision
floating point. Energy is measured in milli-Joules (mJ) required for
the completion of each task. Finally, throughput is measured in Mega-samples
of results produced per second (MSamples/sec) by each kernel.

\subsection{Resource--Precision Performance of GEMM and CONV kernels}

Considering GEMM, out of several sets of experiments performed, we
present results for subblocks with outer dimension of $N=144$, which
corresponds to (or is a multiple of) the setting of other GEMM subblock
kernels (e.g., Eigen%
\footnote{\href{http://eigen.tuxfamily.org/}{http://eigen.tuxfamily.org/} (Eigen
C++ template library)%
}, Goto BLAS \cite{68} and throughput--precision GEMM scaling based
on companding and packing \cite{6}). We then selected two sizes for
the inner dimension of GEMM: $40$ (leading to $144\times40$ by $40\times144$
GEMM subblocks) and $144$ (leading to $144\times144$ by $144\times144$
GEMM subblocks), which represent different operational complexities
for the GEMM subblock realization. Finally, we utilized $L=8$ projections
with coefficients derived via the DCT-II coefficient matrix of \eqref{eq:DCT definition}.

\begin{figure*}[tbh]
\begin{centering}
\includegraphics[scale=0.52]{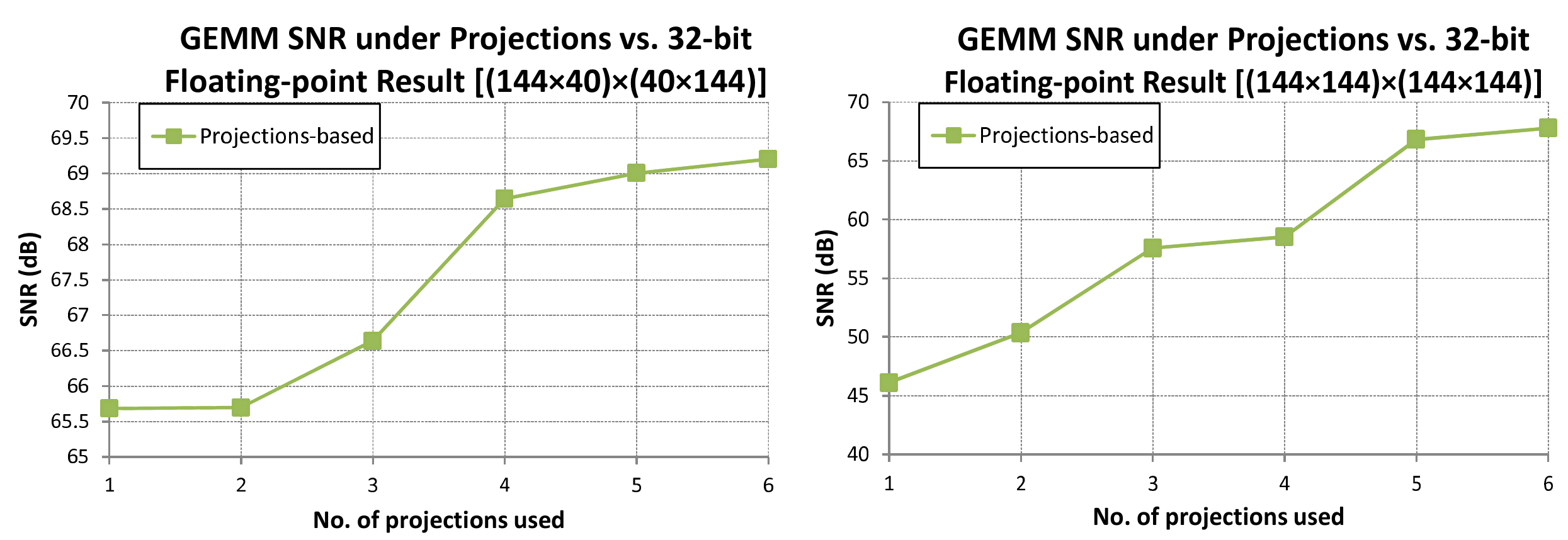} 
\par\end{centering}

\caption{Precision for small and medium-size GEMM inner-dimension (left and
right, respectively). \label{fig:GEMM-PET-results-precision}}
\end{figure*}

\begin{figure*}[tbh]
\begin{centering}
\includegraphics[scale=0.52]{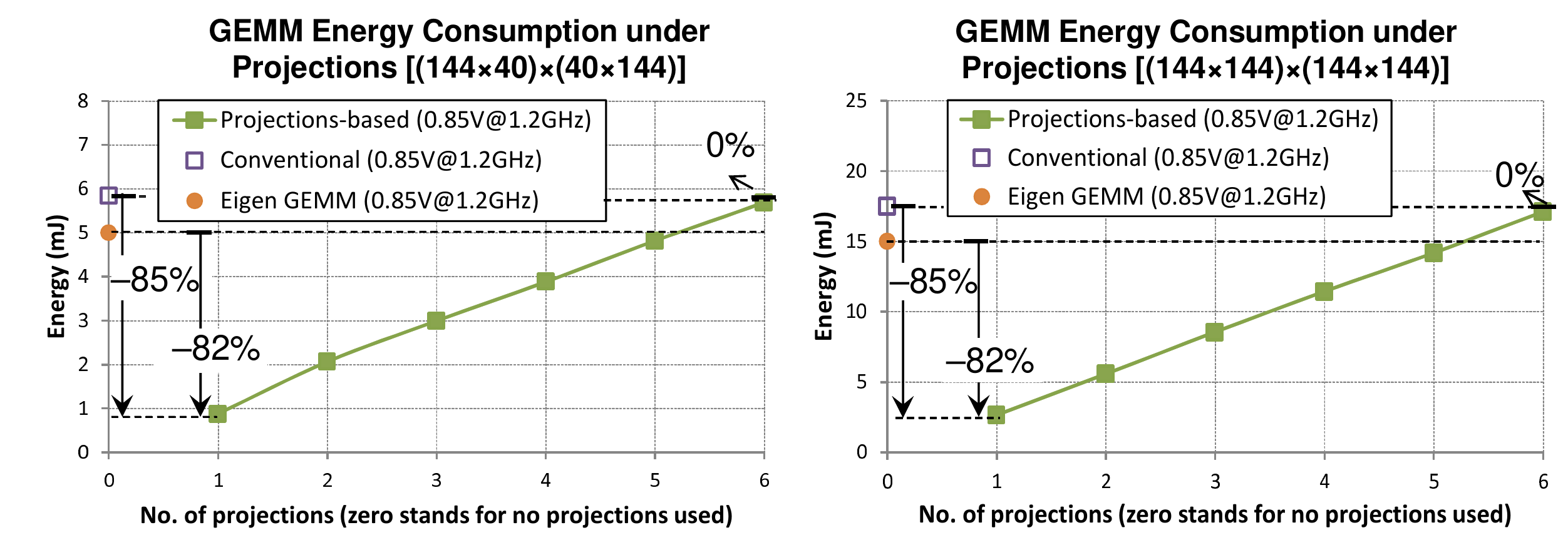} 
\par\end{centering}

\caption{Energy-throughput results for small and medium-size GEMM inner-dimension
(left and right, respectively) under high voltage and high frequency
settings. ``Conventional'' refers to our conventional GEMM realization
that does not utilize projections and it is used as a benchmark. \label{fig:GEMM-PET-results-energy}}
\end{figure*}

\begin{figure*}[tbh]
\begin{centering}
\includegraphics[scale=0.56]{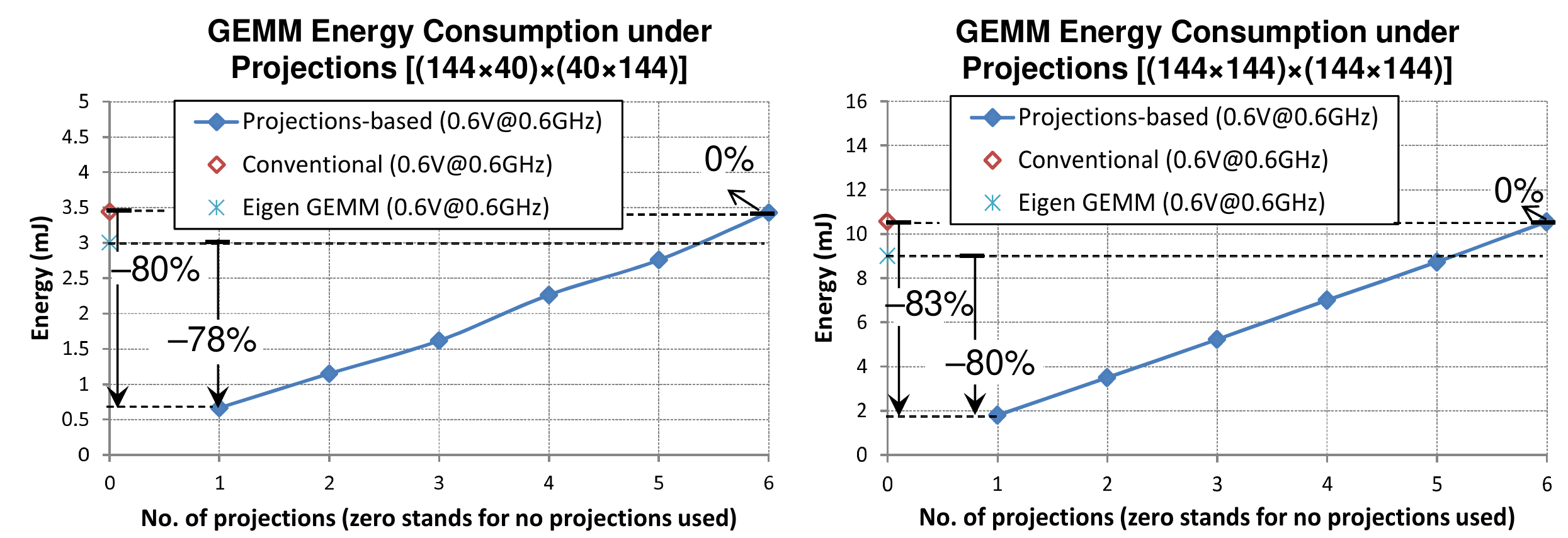} 
\par\end{centering}

\caption{Energy-throughput results for small and medium-size GEMM inner-dimension
(left and right, respectively) under low voltage and low frequency
settings. ``Conventional'' refers to our conventional GEMM realization
that does not utilize projections and it is used as a benchmark. \label{fig:GEMM-PET-results-energy-1}}
\end{figure*}

\begin{figure*}[tbh]
\begin{centering}
\includegraphics[scale=0.52]{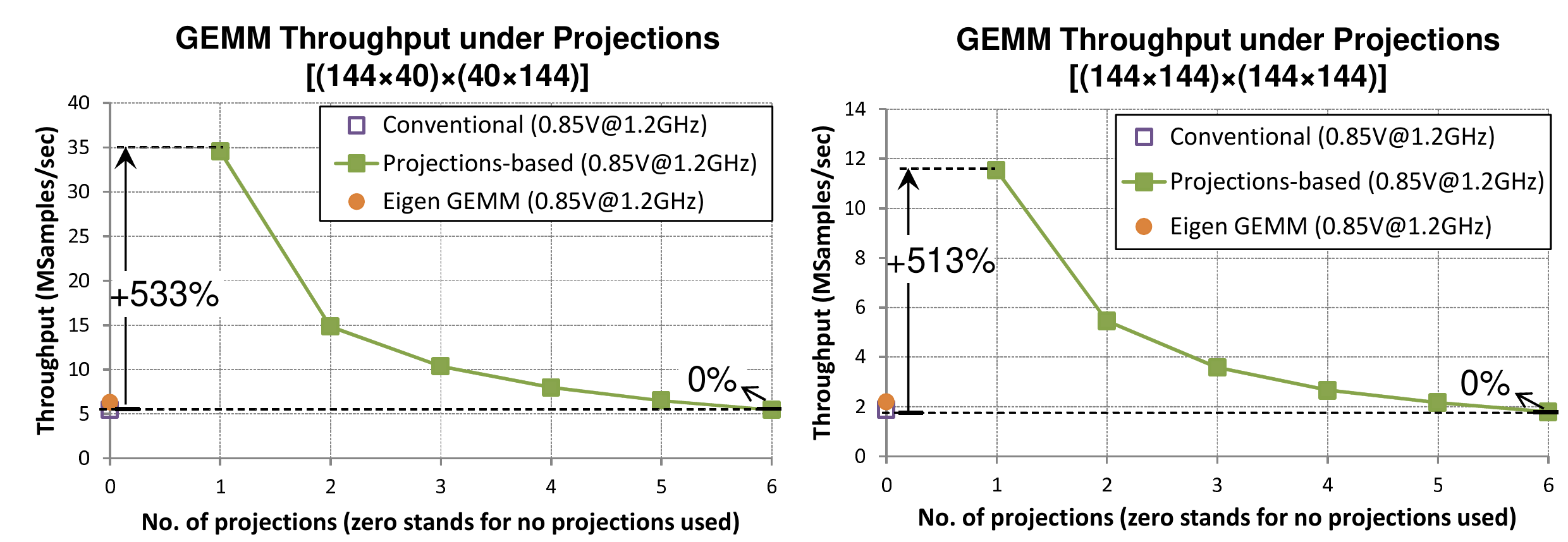} 
\par\end{centering}

\caption{Throughput results for small and medium-size GEMM inner-dimension
(left and right, respectively) under high voltage and high frequency
settings. ``Conventional'' refers to our conventional GEMM realization
that does not utilize projections and it is used as a benchmark. \label{fig:GEMM-PET-results-throughput}}
\end{figure*}

\begin{figure*}[tbh]
\begin{centering}
\includegraphics[scale=0.52]{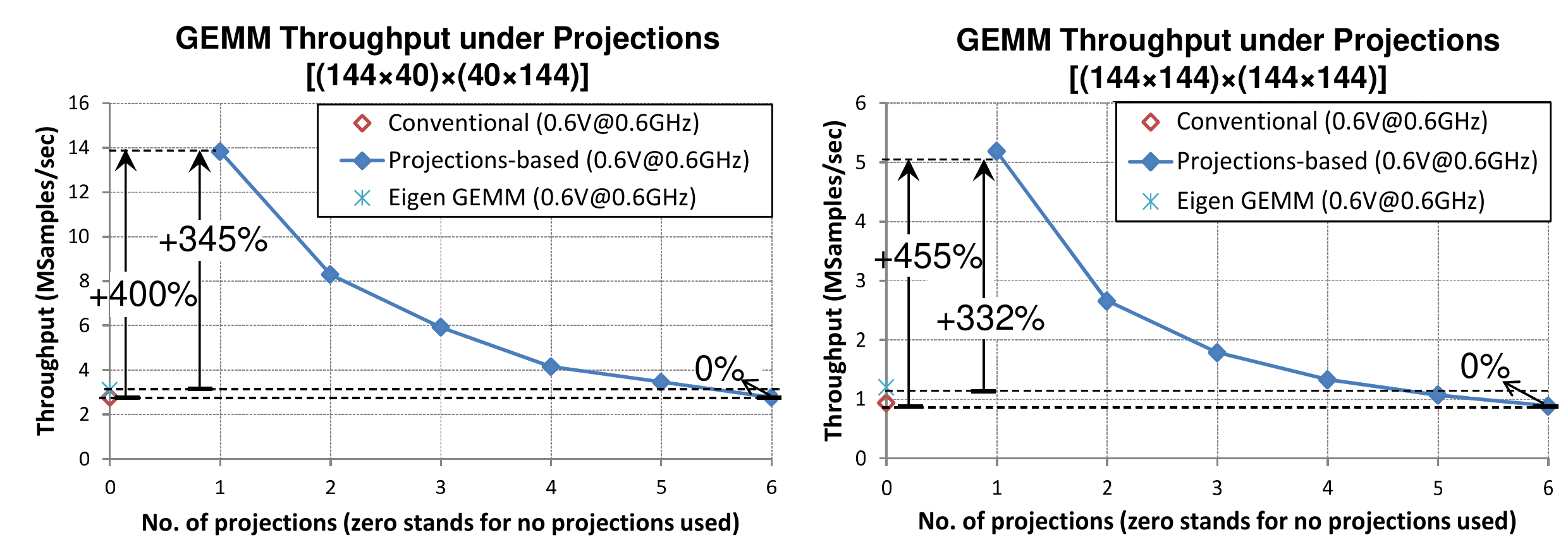} 
\par\end{centering}

\caption{Throughput results for small and medium-size GEMM inner-dimension
(left and right, respectively) under low voltage and low frequency
settings. ``Conventional'' refers to our conventional GEMM realization
that does not utilize projections and it is used as a benchmark. \label{fig:GEMM-PET-results-throughput-1}}
\end{figure*}

Figure \ref{fig:GEMM-PET-results-precision}--Figure \ref{fig:GEMM-PET-results-throughput-1}
present results for precision--energy--throughput scaling against
the conventional GEMM kernel realization, i.e. our SIMD-based GEMM
kernel without projections. Two voltage and frequency levels are used
and, as an external benchmark, we also present results with the Eigen
GEMM kernel%
\footnote{Comparing the energy and throughput efficiency of our own conventional
GEMM realization with the figures obtained with Eigen GEMM shows that
our conventional GEMM kernel is a reasonably high-performing kernel
to benchmark our approach with.%
}.

When using six projections (out of eight), the average SNR is $70$dB
against the conventional GEMM kernel. Under the utilized input range
and GEMM inner dimension, this corresponds to mean square error less
than $7\times10^{-4}$ in the GEMM results, which is deemed acceptable
by all multimedia signal processing applications. This comes at no
overhead in both throughput (in MSamples/sec) and energy consumption
in comparison to the conventional GEMM kernel. 

By reducing the number of projections, our approach achieves up to
$85\%$ reduction in energy consumption against the conventional GEMM
kernel (Figure \ref{fig:GEMM-PET-results-energy} and Figure \ref{fig:GEMM-PET-results-energy-1}).
This substantial reduction is energy comes from the reduction of execution
time while maintaining the same level of power usage. More specifically,
the power usage is identical during the GEMM inner-kenel computation
and only increases by about 5\% during the short time interval required
to perform the projection. However, the projection process allows
for the processing throughput to increase by $315\%\sim533\%$ (Figure
\ref{fig:GEMM-PET-results-throughput} and Figure \ref{fig:GEMM-PET-results-throughput-1},
marginally less improvement is obtained against Eigen GEMM). These
very substantial performance improvements come at the cost of decreasing
the SNR to approximately $46\sim65$ dB in comparison to the result
computed under the conventional realization%
\footnote{However, SNR values above $40$ dB can be regarded as adequate for
many signal processing applications \cite{111}. We remark that these
SNR numbers depend on the dataset and the projection coefficients
used. If projection coefficients are derived specifically for the
data via offline training, e.g. based on principal component analysis
\cite{66}, then it is possible to get even higher SNR values using
an even smaller subset of projection coefficients. However, unlike
a general transform like the DCT, such an approach requires offline
training and is biased towards the dataset selected for the training.
For these reasons, such an exploration is beyond the scope of the
current paper.%
}. We shall show in the next section that such SNR values offer sufficient
accuracy for real-world multimedia recognition and matching systems
utilizing GEMM computations. 

As a final comparison, we evaluated these performance results against
results obtained via throughput--precision GEMM based on our prior
work on companding and packing \cite{6}. On the same hardware platform,
benchmarking the proposed approach under one projection against companding
and packing GEMM led to: \emph{(i)} more than $8$ dB gain in SNR;
\emph{(ii)} more than $60\%$ reduction of energy consumption; and
\emph{(iii)} more than $400$\% increase of throughput. In conjunction
with the fact that any number of projections can be deployed and that
one is free to select projection coefficients suitable to the input
data characteristics, this makes the proposed approach attain significantly
broader resource--precision scalability in comparison to our previously-proposed
companding-and-packing based GEMM. 


Concerning the CONV kernel, we experimented with: block size of $W=20000$
samples, several kernel sizes between $N\in\left[600,1200\right]$
samples, $L=2$ projections using the Haar decomposition coefficients
\cite{112} and producing one projection at half sampling rate (and
interpolating the missing samples) or one projection at full sampling
rate. Representative precision--energy--throughput results with two
settings for the kernel size are given in Tables \ref{tab:Precision-combined}.
We also present comparisons with convolution based on packed processing
\cite{7}, as well as the equivalent results obtained by the conventional
realization of our own SIMD-based CONV kernel (i.e. without projections)
and the CONV kernel of the Cortex-A DSP library commercialized for
ARM Neon by DSP Concepts LLC%
\footnote{\href{http://www.dspconcepts.com/}{http://www.dspconcepts.com/}%
}. The high energy consumption and low throughput reported in Table~\ref{tab:Precision-combined}
for all approaches is due to the large block size used ($20000$ samples).
The results demonstrate that the proposed approach substantially outperforms
packed processing in terms of energy and throughput efficiency, while
allowing for significantly higher SNR. Moreover, it allows for $82\%\sim84\%$
reduction of energy consumption and $360\%\sim400\%$ increase of
processing throughput against the conventional CONV realization. Finally,
while the SNR values of the proposed PET scaling within CONV remain
significantly smaller than the ones of the conventional CONV kernel,
it will be shown in the next section that this does not affect the
accuracy of a real-world application performing audio matching based
on cross-correlation. 

\begin{table*}[tbh]
\caption{Precision, energy and throughput scaling for CONV for two voltage
and frequency levels. The results correspond to average SNR (in dB)
over several inputs, with each SNR value measured against the equivalent
result computed via the conventional CONV kernel. \label{tab:Precision-combined}}

\centering{}%
\begin{tabular}{|c||c|c|c|c|c|c|c|c|c|c|}
\hline 
Method & \multicolumn{2}{c|}{Precision} & \multicolumn{4}{c|}{Energy (mJ)} & \multicolumn{4}{c|}{Throughput (MSamples/sec)}\tabularnewline
 & \multicolumn{2}{c|}{(dB)} & \multicolumn{2}{c|}{$0.85$V@$1.2$GHz} & \multicolumn{2}{c|}{$0.6$V@$0.6$GHz} & \multicolumn{2}{c|}{$0.85$V@$1.2$GHz} & \multicolumn{2}{c|}{$0.6$V@$0.6$GHz}\tabularnewline
\hline 
Kernel size & $600$ & $1200$ & $600$ & $1200$ & $600$ & $1200$ & $600$ & $1200$ & $600$ & $1200$\tabularnewline
\hline 
\hline 
1 projection, half samples & 19.82 & 22.87 & 596 & 1315 & 429 & 771 & 0.058 & 0.027 & 0.027 & 0.014 \tabularnewline
\hline 
1 projection, all samples & 20.07 & 23.41 & 1258 & 2366 & 826 & 1580 & 0.027 & 0.014 & 0.014 & 0.007\tabularnewline
\hline 
Packed processing \cite{7} & 17.65 & 13.60 & 1235 & 2556 & 850 & 1615 & 0.044 & 0.013 & 0.021 & 0.011 \tabularnewline
\hline 
Conventional CONV & $\infty$ & $\infty$ & 2476 & 4884 & 1654 & 3243 & 0.013 & 0.007 & 0.007 & 0.003\tabularnewline
\hline 
Cortex-A DSP CONV & 141.47 & 143.33 & 2142 & 4005 & 1455 & 2692 & 0.016 & 0.008 & 0.008 & 0.004\tabularnewline
\hline 
\end{tabular}
\end{table*}

\section{Resource--Precision Results within Error-tolerant Multimedia Recognition
and Matching Applications \label{sec:5--}}

The proposed approach can bring important benefits to high-performance
multimedia signal processing systems when the precision of computation
is not of critical importance (error-tolerant systems), or when the
input dataset is intrinsically noisy. This is quite common in image,
video or audio analysis, recognition or matching applications, where
the multimedia samples are contaminated with noise stemming from camera
or microphone sensors or lossy coding systems \cite{111}. Here, we
present two representative applications for the proposed framework
within two well-known image and audio recognition and matching systems
proposed in the literature. While each of the two systems is deployed
for a specific task (i.e. face recognition and music identification),
the underlying algorithms are generic and can be applied to a wide
variety of object recognition and audio matching tasks.

\subsection{Resource--Precision Trade-off in Face Recognition based on Principal
Component Analysis (PCA)}

State-of-the-art techniques for object recognition systems derive
feature matrices and use 2D decomposition schemes via matrix multiplication
in order to match features between a new image and an existing database
of images (e.g. for automatic identification of human faces \cite{66}).
When such deployments run on embedded devices such as smartphones
or smart visual sensors for image analysis and recognition \cite{113},
it is expected that thousands of training and recognition tasks should
be computed with the highest-possible resource--precision capability
of each core in order to minimize the required energy consumption
and maximize the processing throughput. 

Using the proposed approach, one can accelerate the real-time training
and matching process for such applications. Specifically, the accelerated
GEMM via projections can be used for the image covariance scatter
matrix calculation during the training stage, as well as for the feature
extraction from test input images \cite{66}. In the following, we
provide details of such a deployment for the prominent 2D-PCA system
of Yang \emph{et al} \cite{66}, which is widely regarded as one of
the best-performing object recognition algorithms based on principal
components. 

The 2D-PCA algorithm for face recognition comprises three stages:
training, feature extraction and matching. The \emph{training stage}
uses a number of training input images of human subjects and first
calculates the image covariance scatter matrix from $J_{\text{set}}$
zero-mean input images, $\mathbf{A}_{j}$, by: 
\begin{equation}
\mathbf{G}_{j}=\sum_{j=0}^{J_{\text{set}}-1}\mathbf{A}_{j}\mathbf{A}_{j}^{\text{T}}.\label{eq:covariance_matrix-calculation}
\end{equation}
Based on this input training set, it then calculates the projection
matrix comprising a series of projection axes (eigenvectors), 
\begin{equation}
\mathbf{X}=\left[\left.\mathbf{x}_{0}\right|\left.\ldots\right|\mathbf{x}_{D-1}\right],
\end{equation}
with $\mathbf{x}_{i}$, $0\leq i<D$ the orthonormal eigenvectors
of $\mathbf{G}_{j}$ corresponding to its $D$ largest eigenvalues
\cite{66}. Each training-set image is mapped to $\mathbf{X}$ via:
\begin{equation}
\mathbf{Y}_{\text{set},j}=\mathbf{A}_{j}\mathbf{X}.\label{eq:training-projection}
\end{equation}
For the \emph{feature extraction stage}, each new input image, $\mathbf{B}_{i}$
(test image), is mapped to $\mathbf{X}$ via: 
\begin{equation}
\mathbf{Y}_{\text{test},i}=\mathbf{B}_{i}\mathbf{X},\label{eq:test-projection}
\end{equation}
with $\mathbf{Y}_{\text{test},i}$ comprising the feature matrix of
test image $\mathbf{B}_{i}$. Finally, the \emph{matching stage} determines
for each test image the training-set image, $\mathbf{A}_{j_{\mathbf{B}_{i}}^{*}}$,
with the smallest distance in their feature matrices: 
\begin{equation}
j_{\mathbf{B}_{i}}^{*}=\arg\min_{\forall j}\left\Vert \mathbf{Y}_{\text{test},i}-\mathbf{Y}_{\text{set},j}\right\Vert _{F}.\label{eq:test-match}
\end{equation}
The complexity of 2D-PCA is predominantly in the matrix multiplications
required for the construction of $\mathbf{G}_{J}$ of \eqref{eq:covariance_matrix-calculation}
during the training stage and the mapping during the feature extraction,
i.e. $\mathbf{Y}_{\text{test},i}$ of \eqref{eq:test-projection},
as the eigenvalue decomposition required for the creation of $\mathbf{X}$
is only performed once every $J_{\text{set}}$ training images and
very fast algorithms exist for the quick estimation of $j_{\mathbf{B}_{i}}^{*}$
of \eqref{eq:test-match}, such as the matching error measures of
Lin and Tai \cite{114}. 

To examine the impact of projections-based resource--precision scaling
of GEMM, we utilize the proposed approach for all the matrix multiplication
operations of \eqref{eq:covariance_matrix-calculation}, \eqref{eq:training-projection}
and \eqref{eq:test-projection} of 2D-PCA. The Yale-A and Yale-B databases
of face images (\href{http://www.face-rec.org/databases/}{http://www.face-rec.org/databases/})
were used for our experiments and, following prior work \cite{66},
each image was cropped to $288\times288$ pixels (that includes the
face portion) and the mean value was subtracted prior to processing. 

Results from performing all matrix multiplication operations of \eqref{eq:covariance_matrix-calculation},
\eqref{eq:training-projection} and \eqref{eq:test-projection} with
just one out of $L\in\left\{ 8,12,16\right\} $ projections {[}via
the DCT-II coefficients of \eqref{eq:DCT definition}{]} are presented
in Table~\ref{tab:GEMM-face-recognition} for both Yale databases.
Following \cite{66}, the first five images of each of the persons
in each database were used for the training set and the remaining
images per person were used as test images and we set $D=10$. 

Starting with the case of $L=8$ projections, the table demonstrates
that, for all GEMM computations, and under the same recognition accuracy
as the conventional (non projections-based) GEMM, the proposed approach
offers $440\%$ increase in the processing throughput and more than
$80\%$ decrease in energy consumption. If we consider all the other
operations and overheads of the entire face recognition application,
the proposed approach still offers $350\%$ increase in the processing
throughput and $79\%$ decrease in overall energy consumption. Importantly,
we obtain the results of Table~\ref{tab:GEMM-face-recognition} based
on two standard test image libraries (Yale-A and Yale-B databases
of face images) and \emph{without} any algorithmic modification. Instead,
only a simple adjustment of the number of retained projections in
the GEMM computations is required. This is a remarkably straightforward
process compared to the previously-proposed packed processing \cite{6}
that requires resource--precision optimization amongst the subblock
matrix products in order to provide for sufficient precision within
the GEMM operations.

Furthermore, for $L=12$ and $L=16$ projections, Table~\ref{tab:GEMM-face-recognition}
demonstrates that the energy and throughput scaling becomes even more
substantial. Namely, between $672\%\sim850\%$ of increase in throughput
and $88\%\sim91\%$ decrease in energy consumption is obtained against
the conventional GEMM realization, with similar scaling when considering
the entire application. However, these cases incur loss of recognition
accuracy in the application in comparison to the conventional (non
projections-based) GEMM. While this loss of recognition rate appears
to be relatively limited, it can be undesirable in cases where maximizing
the expected recognition rate is of paramount importance. We therefore
conclude that the case of $L=8$ comprises an agreeable operational
point, where substantial performance scaling is offered without any
discernible impact in the application results. 

Given the large performance increase, the lack of apparent degradation
in the average recognition accuracy on both databases can be viewed
as a non-intuitive result. However, this can be explained by the energy
compaction performed by the algorithm itself. Essentially, the projection
compacts the vast majority of the energy of the input images into
one eighth of the data samples (using DCT coefficients) before performing
the matrix product. Since all feature extraction and feature matching
algorithms perform energy compaction anyway (from a large set of pixels
to a few eigenvectors using PCA) in order to remove noise and retain
only the principal components of each image covariance scatter matrix,
the projections-based compaction during the GEMM kernel execution
has limited or no effect on the average recognition accuracy of the
system. 

More broadly, the usage of energy compaction techniques is one of
the primary reasons that error-tolerant multimedia signal analysis,
matching and retrieval systems are known to be robust to noise in
their inputs or intermediate computations \cite{6,57,62,111}. For
instance, concerning multimedia retrieval systems in particular, the
survey of Datta \emph{et al} \cite{57} points to various high-level
analysis and retrieval systems that are robust to noise in the input
data or in the calculated low-level feature points used for matching
and retrieval processes (e.g. corner and edge points in images). Furthermore,
well known studies have already analyzed the resilience of low-level
feature extraction to noise \cite{58} and recent work \cite{59,60,61}
has indicated significant complexity-precision tradeoffs in feature
extraction algorithms by incremental or approximate computation of
their computationally-intensive kernels (transforms, distance metric
calculations, matrix-vector products) in space or frequency domain.
Finally, learning algorithms for large data sets have traditionally
been known to be robust to noise in the input or processed data \cite{62}.
However, as explained in the introduction section, exploiting the
inherent energy compaction properties of error-tolerant multimedia
signal processing and analysis algorithms has only achieved limited
performance scaling in programmable processors \cite{111,30,7} because,
until now, only hardware-oriented approaches \cite{8,9,96,95,94,109,126,110,115}
could scale the precision of computations and achieve significant
energy or throughput scaling.

\begin{table*}[tbh]
\caption{Recognition percentage vs. energy--throughput results for GEMM computations
within the 2D-PCA algorithm for face recognition. All results were
produced with $V_{\text{dd}}=0.6$V at $0.6$GHz. \label{tab:GEMM-face-recognition}}

\centering{}%
\begin{tabular}{|c||c|c|c|c|}
\hline 
\multirow{2}{*}{Method} & Recognition rate ($\%$)  & Recognition rate ($\%$) & Energy per match  & Throughput per match\tabularnewline
 & for Yale-A database & for Yale-B database & (mJ) & (MSamples/sec)\tabularnewline
\hline 
\hline 
Proposed  & \multirow{3}{*}{78.40} & \multirow{3}{*}{86.59} & \multirow{3}{*}{29.99} & \multirow{3}{*}{1.24}\tabularnewline
projections-based &  &  &  & \tabularnewline
GEMM, $L=8$ &  &  &  & \tabularnewline
\hline 
Proposed & \multirow{3}{*}{76.81} & \multirow{3}{*}{83.16} & \multirow{3}{*}{21.42} & \multirow{3}{*}{1.75}\tabularnewline
projections-based &  &  &  & \tabularnewline
GEMM, $L=12$ &  &  &  & \tabularnewline
\hline 
Proposed & \multirow{3}{*}{74.22} & \multirow{3}{*}{80.31} & \multirow{3}{*}{16.44} & \multirow{3}{*}{2.15}\tabularnewline
projections-based &  &  &  & \tabularnewline
GEMM, $L=16$ &  &  &  & \tabularnewline
\hline 
Conventional GEMM & 78.40 & 86.59 & 174.79 & 0.23\tabularnewline
\hline 
Packing-based & \multirow{2}{*}{78.81} & \multirow{2}{*}{86.59} & \multirow{2}{*}{99.58} & \multirow{2}{*}{0.39}\tabularnewline
GEMM \cite{6} &  &  &  & \tabularnewline
\hline 
Eigen GEMM & 78.40 & 86.59 & 141.96 & 0.27\tabularnewline
\hline 
\end{tabular}
\end{table*}

\subsection{Resource--Precision Trade-off in Feature Vector Cross-correlation
within a Music Matching System}

We selected as the second test case a recently-proposed music matching
system that matches cover songs \cite{54} with the songs available
in an existing database. For each input song to be identified, the
system works by extracting beat and tempo data and then matching it
to the (precalculated) beat and tempo database via cross correlation.
Matlab code for this and the sample data were collected from the authors\textquoteright{}
site \cite{54}. Given that this implementation is dominated by the
cross-correlation operations \cite{54}, the only modification performed
was the replacement of the Matlab \texttt{xcorr()} function call with
our CONV kernel running on the ARM test-bed. Thus, in this case each
input block of the cross-correlation corresponds to a song\textquoteright{}s
beat and tempo data and each convolution kernel comprises the beat
and tempo data of a song of the database. The settings used for our
experiments were: average beat rate 120 beats-per-minute, chroma element
central frequency 200Hz \cite{54}.

Concerning our implementation, we utilized one out of $L\in\left\{ 2,4\right\} $
projections and used the Haar decomposition (and synthesis) coefficients.
Table~\ref{tab:CONV-music-matching} demonstrates that these settings
yielded the same matching accuracy for all methods for $L=2$ projections
($53.75\%$ match), while providing up to $286\%$ increase in throughput
(and $75\%$ decrease in energy consumption) in comparison to the
conventional CONV implementation. The overall throughput increase
for the entire music-matching application (i.e., including I/O overhead
and the feature extraction from the original audio) is $273\%$ (and
$72\%$ decrease in energy consumption). The competing acceleration
mechanism, i.e., asymmetric companding-and-packing from our previous
work \cite{7}, turns out to be significantly slower and less energy-efficient
than the proposed approach.

Furthermore, for $L=4$ projections, the matching accuracy of the
proposed approaches decreases ($47.21\%$ match), while providing
for even more substantial throughput and energy scaling in comparison
to the conventional CONV implementation, i.e., $569\%$ and $86\%$
respectively. Nevertheless, the small reduction of the matching accuracy
may make this case undesirable to use in a practical deployment. 

Similarly as for the case of face recognition, the proposed approach
incurs no side effects in the matching accuracy of the system for
$L=2$ projections as the utilized beat and tempo features are inherently
noisy and the retained energy in the feature datasets after the projection
suffices for equally-accurate matching over the test dataset. 

\begin{table}[h]
\caption{Matching accuracy vs. energy-throughput scaling for CONV (cross-correlation)
computations per matching operation within a music identification
application using beat and tempo features. All results were produced
with $V_{\text{dd}}=0.6$V at $0.6$GHz.\label{tab:CONV-music-matching}}

\centering{}%
\begin{tabular}{|c||c|c|c|}
\hline 
\multirow{2}{*}{Method} & \multirow{2}{*}{Matching ($\%$)} & \multirow{2}{*}{Energy (mJ)} & \multicolumn{1}{c|}{Throughput }\tabularnewline
 &  &  & (MSamples/sec)\tabularnewline
\hline 
\hline 
Proposed  & \multirow{3}{*}{53.75} & \multirow{3}{*}{2122} & \multirow{3}{*}{0.027}\tabularnewline
projections-based &  &  & \tabularnewline
CONV, $L=2$ &  &  & \tabularnewline
\hline 
Proposed & \multirow{3}{*}{47.21} & \multirow{3}{*}{1123} & \multirow{3}{*}{0.046}\tabularnewline
projections-based &  &  & \tabularnewline
CONV, $L=4$ &  &  & \tabularnewline
\hline 
Conventional  & \multirow{2}{*}{53.75} & \multirow{2}{*}{8254} & \multirow{2}{*}{0.007}\tabularnewline
CONV &  &  & \tabularnewline
\hline 
Packing-based  & \multirow{2}{*}{53.75} & \multirow{2}{*}{4284} & \multirow{2}{*}{0.021}\tabularnewline
CONV \cite{7} &  &  & \tabularnewline
\hline 
Cortex-A DSP  & \multirow{2}{*}{53.75} & \multirow{2}{*}{7264} & \multirow{2}{*}{0.008}\tabularnewline
CONV &  &  & \tabularnewline
\hline 
\end{tabular}
\end{table}

\section{Conclusion\label{sec:7}}

We propose an approach to systematically trade-off precision for substantial
energy and throughput scaling in generic matrix multiplication (GEMM)
and discrete convolution (CONV) kernels. Given that our approach applies
linear projections within the top-level processing of these kernels,
it allows for seamless scaling of resources versus the accuracy of
the performed computations without cumbersome and algorithm- or application-specific
customization. Experiments with the recently-introduced ARM Cortex
A15 processor on a dedicated test-bed supporting different voltage
and frequency levels and accurate energy measurement, demonstrate
that our proposal leads to more than five-fold reduction of energy
consumption and more than five-fold increase of processing throughput
against the conventional (i.e., non projections-based) realization
of GEMM and CONV kernels. Experimental results within multimedia recognition
and matching applications show that the precision loss incurred by
the proposed projections-based GEMM and CONV kernels can be tolerated
with limited or no noticeable effect on the recognition and matching
accuracy of applications and that our proposal allows for truly dynamic
adaptation without incurring reconfiguration overheads. 

The proposed approach opens up a new avenue for dynamic precision--energy--throughput
scaling within high-performance GEMM and CONV kernel designs. For
the first time, linear transforms can be used towards dynamic resource
scaling of such kernels with graceful precision degradation. Even
though in this paper we used well-known non-adaptive transforms for
the projection coefficients, such as the discrete cosine transform
and the Haar transform, if training input datasets are available \emph{a-priori},
projections based on principal component analysis could be employed
(with their coefficients derived offline) for optimized precision--energy--throughput
scaling within each error-tolerant multimedia application. Alternatively,
if feedback on the incurred imprecision in the results is available
via the application, the projection mechanism of the GEMM and CONV
kernels can be tuned to \emph{learn} the best projection parameters.
These are aspects that can be explored in future work. 

\bibliographystyle{IEEEbib}

\end{document}